\documentclass[a4paper,11pt]{article}
\usepackage{jcappub} % JCAP/JHEP style
\usepackage{lineno}
\usepackage{color}
\usepackage[dvipsnames]{xcolor}
\usepackage{comment}
\usepackage{float}
\usepackage{orcidlink}
\usepackage{subcaption} % For subfigure environment

\newcommand\jm{\textcolor{blue}}

\title{Bright Siren Cosmology Beyond Fisher: DALI Forecasts for Third-Generation Gravitational Wave Observatories}

\author[a]{M. H. Pacheco\orcidlink{0000-0003-0635-7484},}
\author[b]{J. M. S. de Souza\orcidlink{0000-0003-1552-0095},}
\author[a]{O. D. Aguiar\orcidlink{0000-0002-2139-4390},}
\author[a]{M. E. S. Alves\orcidlink{0000-0002-7063-694X}}

\affiliation[a]{Instituto Nacional de Pesquisas Espaciais, São José dos Campos, SP, Brazil}
\affiliation[b]{Núcleo Cosmo-Ufes \& PPGCosmo, Universidade Federal do Espírito Santo, Vitória, ES, Brazil}

\emailAdd{mayara.pacheco@inpe.br}
\emailAdd{josiel.souza@ufes.br}
\emailAdd{odylio.aguiar@inpe.br}
\emailAdd{marcio.alves@inpe.br}

\abstract{
Gravitational waves (GW) standard sirens provide a calibration-free measurement
of the luminosity distance and constitute an independent probe of the expansion history of the Universe, with
particular relevance for the ongoing Hubble constant tension. Forecasting the cosmological
capacity of future detector networks is commonly performed with Fisher-matrix methods,
which assume locally Gaussian likelihoods and can fail in the presence of strong degeneracies
and non-linear correlations (notably the distance--inclination coupling). In this work we
assess non-Gaussian extensions of Fisher forecasting based on the Derivative Approximation
for Likelihoods (DALI). We simulate ensembles of bright siren catalogs for third-generation
observatories, adopting a Cosmic Explorer (CE) and a triangular Einstein Telescope (ET) network, frequency-domain inspiral waveforms, and
selection-motivated priors for counterpart detection. We reconstruct approximate likelihoods
using Fisher and the first DALI correction (Doublet), and propagate the inferred
distance information to cosmological inference in a $\omega_0\omega_a$CDM model.
We find that the Doublet approximation provides tighter cosmological constraints than
the Fisher approximation, reducing the inferred uncertainties by approximately a factor
of two in agreement with non-approximate likelihood (called Exact) results. For our GW network configuration, 300 bright-siren detections up to \(z=1\) yield an \(H_0\) precision of \(\sim 2.7\%\) with the Doublet approximation, compared with \(\sim 2.5\%\) for the exact likelihood. The adopted non-Gaussian GW priors induce a systematic displacement of the
inferred luminosity distances from the fiducial values, which is also present in the
exact GW reconstruction and propagates into the cosmological parameters. Compared
with the exact cosmological likelihood, the Doublet results differ by only $\sim 10\%$
on average in the parameter uncertainties, compared with $\sim 195\%$ for the
traditional Fisher approximation. The Doublet approximation also substantially reduces
the computational cost, by a factor of $\sim 55$ for the GW analysis and by $\sim 12$
for cosmological likelihood evaluations, yielding an estimated reduction of
$\sim 55$ in the total computational cost of the combined analysis.
Our results show the potential of DALI-based approximations to provide accurate and computationally efficient tools for forecasting in bright siren cosmology.
}

\begin{document}

\maketitle
\flushbottom

\section{Introduction}
\label{sec:intro}

\hspace{1.0cm} Gravitational wave (GW) observations have opened an observational window onto the Universe, providing direct access to compact-binary coalescences and enabling distance measurement methods that differ from those obtained through electromagnetic probes. In the standard-siren framework, originally introduced by Schutz \cite{Schutz1986}, compact-binary coalescences can be used as distance indicators because the GW signal directly encodes the luminosity distance $d_L$ to the source, without the need for external calibration \cite{HolzHughes2005,Fishbach2019}. At leading order, the amplitude of the inspiral waveform scales inversely with the $d_L$, but it also depends on the redshifted chirp mass $\mathcal{M}_c$ and on the binary orientation, particularly the inclination angle $\iota$. Therefore, the distance information is not obtained from the amplitude alone.

The time-derivative of the inspiral frequency $\dot{f}$ depends on the $\mathcal{M}_c$ with the same functional dependence as the waveform amplitude. Thus, one concludes that $d_L$ can be determined directly from GW data $h(t)$ and $\dot{f}(t)$. In this sense, compact-binary coalescences, like binaries of black holes (BBH), neutron stars (BNS) or mixed neutron star-black holes (NSBH), behave as absolute distance indicators \cite{Schutz1986}. In practice, however, the inference of $d_L$ is limited by parameter degeneracies, most notably the $d_L-\iota$ degeneracy, which can generate asymmetric and non-Gaussian posteriors for individual events.

GW observations alone do not directly provide the source redshift $z$, since the waveform depends on detector-frame masses rather than on intrinsic masses and redshift separately. For cosmological applications, an additional redshift measurement is therefore required. In this work, we focus on bright sirens, namely GW events with an identified electromagnetic counterpart. In such cases, the counterpart allows the host galaxy to be identified and its redshift to be measured spectroscopically, so that each event provides a direct point in the $d_L-z$ relationship. Once both $d_{L}$ and $z$ are available, the source provides a direct and calibration-independent measurement of the cosmic expansion history \cite{HolzHughes2005, Fishbach2019}. In the low-redshift regime, this relation enables an inference of the Hubble constant $H_{0}$.

The binary neutron-star merger GW170817 provided the first GW standard siren measurement of the Hubble constant \cite{abbott2017}. Electromagnetic observations identified an optical transient spatially associated with the galaxy NGC~4993, allowing the host-galaxy redshift to be measured independently of the GW signal. By combining $d_L$ inferred from the GW data with the recession velocity inferred from electromagnetic observations, it was obtained $H_{0}=70.0^{+12.0}_{-8.0}\ \mathrm{km\,s^{-1}\,Mpc^{-1}}$\cite{abbott2017}.
 
 It has been shown that as the number of detections increases, our constraints on $H_0$ can become more precise, see \cite{Chen2018, Feeney2019, vitale2018}. Chen et al.~\cite{Chen2018}, for instance, evaluated the ability of second-generation GW detectors to constrain the $H_0$ using both BNS and BBH mergers. Redshift estimation was based either on electromagnetic counterparts, as in GW170817, or on the redshift distribution of host galaxies or galaxy clusters. Their results indicate that observations of approximately 10, 60, and 200 BNS mergers with electromagnetic counterparts could constrain $H_0$ to precisions of $4\%$, $2\%$, and $1\%$, respectively. Unfortunately, this prediction is limited by the small number of observed BNS mergers \cite{LVK2025GWTC4}. 

Third-generation GW observatories (3G), like Einstein Telescope (ET) and Cosmic Explorer (CE), will provide a substantial advance over current second-generation detectors, particularly in two aspects: the sky localization of nearby compact-binary sources and the precision for which $d_L$ can be inferred \cite{Punturo:2010zz,Reitze:2019iox,zhao2026}. Better localization increases the likelihood of associating GW events with electromagnetic counterparts, while more accurate distance measurements reduce one of the dominant uncertainties in standard siren estimates of $H_0$. Zhao et al.~\cite{zhao2018} investigated different configurations of 3G detector networks and evaluated their ability to constrain cosmological parameters using BNS and NSBH mergers. They found that a network of ET-like detectors distributed across the United States, Europe, and Australia could measure the $d_L$ of a BNS merger at $z = 0.1$ with a fractional uncertainty ranging from about $10\%$ to $1\%$. For a sample of 100 comparable events with identified electromagnetic counterparts at that redshift, the resulting measurement of the $H_0$ could reach a precision of approximately $0.3\%$. 

This expectation is supported by several independent forecasts. Taylor and Gair \cite{TaylorGair2012} found that ET could catalog tens to hundreds of thousands of BNSs per year, and that a sample of $10^5$ detections could constrain dark-energy equation-of-state parameters with an accuracy comparable to future CMB+BAO+SNIa measurements. Similarly, BBHs forecasts for 3G networks indicate that an ET+CE+CE configuration, combined with a galaxy catalog information complete up to $z=1$, can constrain $H_0$ to about $0.8\%$ and $\Omega_m$ to about $10\%$ within one year of observations. On the other hand, bright siren forecasts with future ET and CE observations further indicate that BNS and NSBH events with electromagnetic counterparts can provide independent constraints on the redshift evolution of the dark-energy equation of state \citep{Afroz2025,koehn2026binary}. Together, these forecasts indicate that with 3G detectors, GW astronomy will become a precise probe of the late-time expansion of the Universe.

 However, robust forecasting in this context requires statistical methods that remain reliable when the likelihood departs from Gaussianity. For this reason, Fisher-matrix (FM) methods, although computationally efficient and widely used \cite{vallisneri2008, gwbench, gwfish, gair2023, gwfast}, are not always accurate when the likelihood exhibits strong curvature, skewness, or non-linear correlations \cite{gwdali, dali2019, cosmodali, Rodriguez2013, satej2013, Wolz2012}. Higher-order likelihood expansions provide an extension of the Fisher approach while retaining its derivative-based structure. In this work, we investigate this problem using the Derivative Approximation for Likelihoods (DALI) formalism \cite{sellentin2014, dali2019, gwdali, cosmodali}. We compare the Gaussian Fisher approximation with higher-order DALI expansions in their ability to reconstruct the likelihood of bright standard sirens and to propagate the inferred distance information to cosmological parameter estimation.

Although standard sirens can also be used statistically in the absence of an electromagnetic counterpart, such dark-siren approaches are not the focus of the present analysis. Here we restrict attention to sources with electromagnetic counterparts, for which the redshift information is directly available, and the central statistical challenge relies on understanding how non-Gaussian structures in the GW likelihood propagate into cosmological-parameter inference. Therefore, bright sirens provide the setting adopted in this work to assess the performance of higher-order likelihood expansions beyond the Fisher approximation.

Our analysis is based on simulated bright siren catalogs for 3G networks. We assess how accurately Fisher and DALI approximations reproduce the underlying likelihood structure, with particular attention to the $d_L-\iota$ degeneracy, and we examine the consequences of these approximations for cosmological inference. Our goal is to determine under which conditions the Gaussian approximation remains adequate and when higher-order DALI corrections become necessary for reliable forecasting with future GW observations.

Recent studies of the ET science case have investigated the cosmological potential of bright standard sirens under different assumptions about detector configurations and electromagnetic follow-up strategies. For a 10-km triangular ET configuration, forecasts report marginalized uncertainties on the Hubble constant of approximately $0.9\%$ for a one-year gravitational-wave catalog with kilonova counterparts and $5.7\%$ for a five-year catalog with gamma-ray burst counterparts~\cite{Branchesi2023,ETBlueBook2025}. These results illustrate how the expected cosmological precision depends on the assumed source population, detector configuration, and electromagnetic follow-up strategy. In this context, rather than providing an updated end-to-end forecast for the ET science case, the present work investigates how non-Gaussian gravitational-wave likelihood approximations affect cosmological inference for a fixed bright-siren catalog, focusing on the accuracy of Fisher and DALI-based methods.

The remainder of this paper is organized as follows. Section~2 reviews the Fisher matrix and DALI formalisms, highlighting their underlying assumptions and limitations. Section~3 describes the construction of the simulated gravitational-wave catalogs, while Section~4 presents the gravitational-wave parameter-estimation methodology, including the Fisher and DALI approximations. Section~5 presents the resulting cosmological constraints and provides a quantitative comparison of the different approximations. Finally, Section~6 summarizes our main findings and discusses prospects for future work.

\section{Likelihood Approximations}

\hspace{1.0cm}  Throughout this work, we refer to the exact likelihood as the unapproximated functional form used for the inference, namely the standard Gaussian-noise likelihood in the GW analysis and the \(\chi^2\)-based likelihood in the cosmological analysis.
We also work with geometrized units, where $G=c=1$.

\subsection{Fisher matrix formalism}
\label{sec:method}

\hspace{1.0cm}  The Fisher information matrix provides a convenient way to predict the expected uncertainties on model parameters before any data are collected. For a model with \(\mathcal{N}\) parameters, it typically requires only \(\mathcal{O}(\mathcal{N})\) likelihood or model evaluations \cite{cosmodali}. This scaling should be understood as follows: the Fisher matrix is constructed from derivatives of the theoretical predictions with respect to each parameter, evaluated around a fiducial point \cite{tegmark97, Tegmark1997}. In a finite-difference implementation, one evaluates the model at the fiducial point and at a fixed number of displaced points along each of the \(\mathcal{N}\) parameter directions. For instance, a central-difference scheme requires approximately \(1+2\mathcal{N}\) evaluations. Since constant factors are ignored in asymptotic notation, this cost scales linearly with the number of parameters, i.e. as \(\mathcal{O}(\mathcal{N})\). Once these derivative vectors are obtained, the individual Fisher-matrix elements are computed through algebraic contractions, without requiring a full exploration of the likelihood surface.

The method follows from a Taylor expansion of the log-likelihood around its maximum. The derivatives of the theoretical predictions with respect to the parameters quantify how sensitive the observables are to small changes in the parameter vector \(\boldsymbol{\theta}\). If the likelihood \(\mathcal{L}(\boldsymbol{\theta})\) is well approximated by a Gaussian in the neighborhood of its peak, then \(\mathcal{L}\) is locally quadratic and the parameter constraints are determined by the curvature of \(log\mathcal{L}\) \cite{tegmark97, Tegmark1997}, which can be written as
\begin{equation}
\mathcal{L}(\boldsymbol{\theta})
\propto
\exp\left[
-\frac{1}{2}
(\Delta\boldsymbol{\theta})^{T}
\mathbf{F}
(\Delta\boldsymbol{\theta})
\right],
\label{eq:fisher_likelihood}
\end{equation}
where $\mathbf{F}$ is the Fisher information matrix and $\Delta \boldsymbol{\theta}$ represents deviations from the best-fit parameter values. In this approximation, the inverse of the Fisher matrix corresponds to the covariance matrix of the estimated parameters,
\begin{equation}
\mathbf{C} = \mathbf{F}^{-1},
\end{equation}
such that
\begin{equation}
\sigma_{\theta_i} = \sqrt{C_{ii}}
\end{equation}
gives the minimum expected uncertainty for each parameter. This result is known as the Cram\'er--Rao bound.

In the GW data analysis context, codes such as \texttt{GWBENCH} \cite{gwbench}, \texttt{GWFAST} \cite{gwfast} and \texttt{GWFISH} \cite{gwfish} implement this approach to rapidly evaluate the sensitivity of proposed observatories (e.g., ET, CE, LISA) or multi-band networks. Similarly, in cosmology, the Fisher information matrix is widely used to forecast constraints on parameters such as the matter density parameter $\Omega_M$, the amplitude of matter fluctuations $\sigma_8$, the dark-energy equation-of-state parameter $\omega_0$, or the sum of neutrino masses \cite{More2013}. 

Nevertheless, Fisher forecasting has important limitations. The Fisher matrix method provides an exact description only when the posterior is Gaussian in the parameters; consequently, it predicts symmetric, elliptical constraint contours. In practice, this approximation can break down, particularly in the presence of strong parameter degeneracies or when the data are only weakly informative, leading to inaccurate uncertainty estimates \cite{cosmodali}.

% <<<<<<<<<<<<<<<<<<<<<<<<<<<<<<<<<<<<<<<<<<<<<<<<<<<<<<<<<<<<<<<<<<<<<<<<<<<<<<<<<<<<<<<<<<<<<<<<<<<<<<<<<<<<<<<<<<<<<<<<<<<
\subsection{Derivative Approximation for Likelihoods (DALI)}
\label{sec:dali_method}

\hspace{1.0cm} If the posterior is non-Gaussian, the Fisher approximation can become inaccurate, particularly when the observables depend nonlinearly on the model parameters, when the likelihood displays curved degeneracy directions, or when weakly constraining data produce non-elliptical posterior contours \cite{Rodriguez2013,Wolz2012,Bellomo2020,cosmodali,dalichina}. To overcome these limitations, the Derivative Approximation for Likelihoods (DALI) extends the Fisher formalism by incorporating higher-order derivatives of the likelihood or, equivalently, of the theoretical model, while preserving a positive definite and normalizable approximation to the posterior \cite{sellentin2014,sellentin2015,cosmodali,dalichina, dali2019}. In its lowest non-Gaussian correction, known as the Doublet approximation, the method requires only \(\mathcal{O}(\mathcal{N}^{2})\) likelihood or model evaluations for \(\mathcal{N}\) parameters \cite{sellentin2014,cosmodali}.

This scaling arises because the Doublet approximation depends not only on the first derivatives of the theoretical observables, \(\boldsymbol{\mu}_{,a}\), but also on their second derivatives, \(\boldsymbol{\mu}_{,ab}\). These second derivatives must be evaluated for all independent pairs of parameters \((a,b)\). For a model with \(\mathcal{N}\) parameters, the number of such independent pairs is
\begin{equation}
N_{\rm pair} = \frac{\mathcal{N}(\mathcal{N}+1)}{2},
\label{eq:number_parameter_pairs_dali}
\end{equation}
which scales quadratically with the number of parameters. In a finite-difference implementation, each derivative is obtained from a fixed number of model evaluations around the best-fit point. Hence, although the numerical prefactor depends on the derivative stencil adopted, the leading computational cost scales as
\begin{equation}
N_{\rm eval}^{\rm Doublet} \propto \frac{\mathcal{N}(\mathcal{N}+1)}{2} = \mathcal{O}(\mathcal{N}^{2}).
\label{eq:dali_on2_scaling}
\end{equation}

Thus, Doublet is more expensive than the Fisher approximation, whose leading cost is associated with \(\mathcal{O}(\mathcal{N})\) first-derivative evaluations, but it remains much less costly than a full sampling of the exact likelihood \cite{cosmodali,sellentin2014,sellentin2023}.

Following the same local-expansion logic underlying Fisher forecasting, let us define
\begin{equation}
\Delta\theta^a \equiv \theta^a-\hat{\theta}^{a},
\label{eq:dali_delta_theta}
\end{equation}
where \(\hat{\theta}^{a}\) denotes the best-fit. Since the first derivatives vanish at the maximum, the Fisher approximation corresponds to retaining only the quadratic term of the expansion,
\begin{equation}
\mathcal{L}(\boldsymbol{\theta}) \simeq \mathcal{N} \exp\!\left[ -\frac{1}{2} F_{ab}\Delta\theta^a\Delta\theta^b \right],
\label{eq:fisher_approx_dali_section}
\end{equation}
with
\begin{equation}
F_{ab} 
=
-\left. \frac{\partial^2 \log\mathcal{L}} {\partial\theta^a\partial\theta^b} \right|_{\boldsymbol{\hat{\theta}}}.
\label{eq:fisher_tensor_loglikelihood}
\end{equation}

On the other hand, DALI reorganizes the expansion not only by powers of \(\Delta\theta^a\), but also by the order of derivatives of the model predictions \(\boldsymbol{\mu}\) \cite{sellentin2015}. For a likelihood given by: 
\begin{equation}
    \log\mathcal{L} = -\frac{1}{2}\boldsymbol{(d-\mu)}^T\boldsymbol{M(d-\mu)}    
\end{equation} 
where $\mathbf{d}$ is the vector of data and $\boldsymbol{M}$ encodes the uncertainties of the measurements, the second-order derivative expansion, known as Doublet, is
\begin{equation}
\begin{aligned}
\mathcal{L}_{\rm Doublet} (\boldsymbol{\theta})
&=
\mathcal{N} \exp\Bigg[ -\frac{1}{2} \boldsymbol{\mu}_{,a}^{T} \mathbf{M} \boldsymbol{\mu}_{,b} \Delta\theta^{ab}
-\frac{1}{2} \boldsymbol{\mu}_{,ab}^{T} \mathbf{M} \boldsymbol{\mu}_{,c} \Delta\theta^{abc}
-\frac{1}{8} \boldsymbol{\mu}_{,ab}^{T} \mathbf{M} \boldsymbol{\mu}_{,cd} \Delta\theta^{abcd} \Bigg],
\end{aligned}
\label{eq:doublet_dali_derivative_expansion}
\end{equation}
 where commas denote partial derivatives and multiple upper indices in $\Delta\theta$ denotes multiples products, e.g. $\Delta\theta^{ab}\equiv\Delta\theta^a\Delta\theta^b$, $\Delta\theta^{abc}\equiv\Delta\theta^a\Delta\theta^b\Delta\theta^c$ and so on.
 
The Triplet approximation extends this construction by including third derivatives of the model predictions:
\begin{equation}
\begin{aligned}
\mathcal{L}_{\rm Triplet} (\boldsymbol{\theta})
&=
\mathcal{N} \exp\Bigg[ -\frac{1}{2} \boldsymbol{\mu}_{,a}^{T} \mathbf{M} \boldsymbol{\mu}_{,b} \Delta\theta^{ab}
\\
&\hspace{1.2cm}
-\Bigg(\frac{1}{2} \boldsymbol{\mu}_{,ab}^{T} \mathbf{M} \boldsymbol{\mu}_{,c} \Delta\theta^{abc}
+
\frac{1}{8} \boldsymbol{\mu}_{,ab}^{T} \mathbf{M} \boldsymbol{\mu}_{,cd} \Delta\theta^{abvd} \Bigg)
\\
&\hspace{1.2cm}
-\Bigg(\frac{1}{6} \boldsymbol{\mu}_{,a}^{T} \mathbf{M} \boldsymbol{\mu}_{,bcd} \Delta\theta^{abcd}
+
\frac{1}{12} \boldsymbol{\mu}_{,abc}^{T} \mathbf{M} \boldsymbol{\mu}_{,de} \Delta\theta^{abcde}
+ \frac{1}{72} \boldsymbol{\mu}_{,abc}^{T} \mathbf{M} \boldsymbol{\mu}_{,def} \Delta\theta^{abcdef} \Bigg) \Bigg].
\end{aligned}
\label{eq:triplet_dali_derivative_expansion}
\end{equation}

\begin{figure}[htbp]
  \centering
  \includegraphics[width=.7\linewidth]{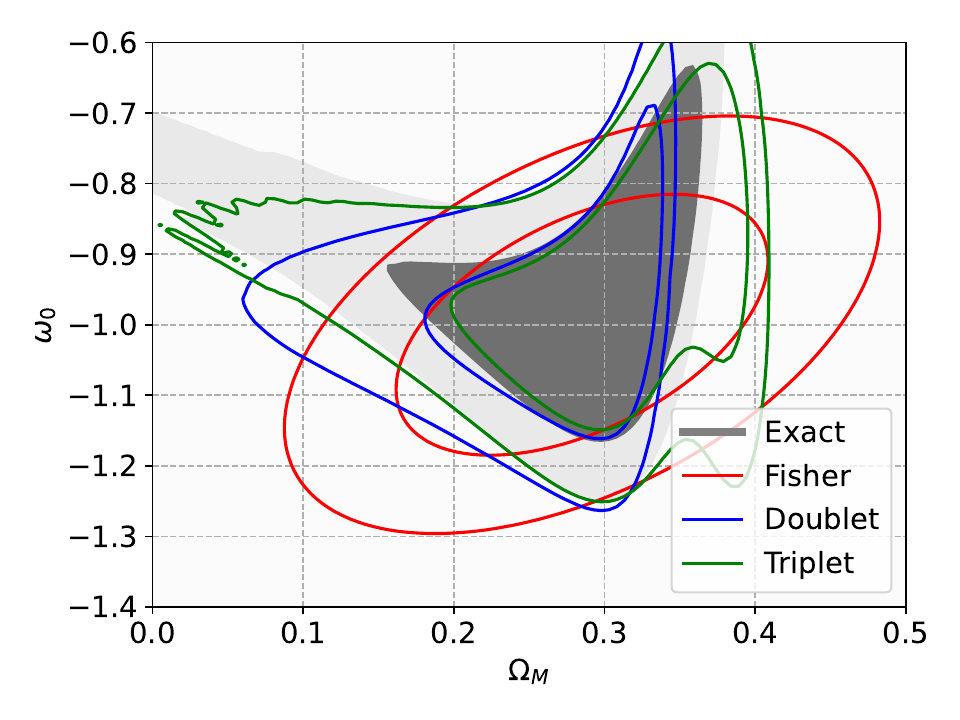}
    \caption{Two-dimensional constraints in the $(\Omega_M, \omega_0)$ plane obtained with three supernovae likelihood approximations: Fisher (red), Doublet (blue), and Triplet (green) following~\cite{sellentin2014}. The dark-gray and gray regions shows the exact posterior density with contours indicating the $68\%$ and $95\%$ credible regions ($1\sigma$ and $2\sigma$). The comparison highlights how the inclusion of higher-order DALI terms captures non-Gaussian structure and parameter degeneracies beyond the Fisher approximation.}
    \label{fig:DALI_initial}
\end{figure}

Thus, Fisher contains only the first derivatives of the model and produces a Gaussian approximation in parameter space; Doublet includes second derivatives, \(\boldsymbol{\mu}_{,ab}\); and Triplet includes third derivatives, \(\boldsymbol{\mu}_{,abc}\). The method reduces to the Fisher limit when the posterior is effectively Gaussian, but it can capture curved, skewed, or otherwise non-elliptical posterior contours when the model depends nonlinearly on its parameters (see Fig.~\ref{fig:DALI_initial}) \cite{sellentin2014,sellentin2015,cosmodali,dalichina,sellentin2023}.

In GW data analysis, the data vector is replaced by the measured strain \(g(t)\), and the theoretical prediction is the waveform \(h(t;\boldsymbol{\theta})\). Assuming stationary Gaussian noise with one-sided power spectral density \(S_n(f)\), the likelihood can be written as
\begin{equation}
\mathcal{L}_{\rm GW} \left(g\mid\boldsymbol{\theta}\right) \propto \exp\!\left[-\frac{1}{2} \left\langle g-h(\boldsymbol{\theta}) \,\middle|\, g-h(\boldsymbol{\theta}) \right\rangle \right],
\label{eq:gw_gaussian_likelihood}
\end{equation}
where the noise-weighted inner product is defined as
\begin{equation}
\left\langle a\mid b\right\rangle = 4\,{\rm Re} \sum_{i=1}^{N_{dets}}\int_{f_{\rm min}}^{f_{\rm max}} \frac{\tilde{a}_i^{\,*}(f)\tilde{b}_i(f) } {S_{n,i}(f)} \,df .
\label{eq:gw_inner_product}
\end{equation}
where $N_{dets}$ is the number of GW detectors.

With the replacement
\begin{equation}
\boldsymbol{\mu}_{,a}^{T} \mathbf{M} \boldsymbol{\mu}_{,b} \longrightarrow \left\langle h_{,a}\mid h_{,b}\right\rangle,
\label{eq:mu_to_waveform_replacement}
\end{equation}
the Fisher matrix approach becomes
\begin{equation}
    \log\mathcal{L}_{GW}^{Fisher} = \log \mathcal{L}_{\boldsymbol{\hat{\theta}}} -\frac{1}{2} \sum_{a,b} \left\langle h_{,a} \mid h_{,b} \right\rangle \Delta\theta^{ab}
\end{equation}

For the Doublet approximation we have~\cite{dalichina}
\begin{eqnarray}
\log \mathcal{L}_{\rm GW}^{\rm Doublet}
&=&
\log\mathcal{L}_{GW}^{Fisher} -\frac{1}{2} \sum_{a,b,c} \left\langle h_{,a} \mid h_{,bc} \right\rangle \Delta\theta^{abc}
 -\frac{1}{8} \sum_{a,b,c,d} \left\langle h_{,ab} \mid h_{,cd} \right\rangle \Delta\theta^{abcd} .
\label{eq:gw_doublet_dali_loglikelihood}
\end{eqnarray}

Finally, the Triplet approximation for GW becomes~\cite{dalichina}
\begin{eqnarray}
\log \mathcal{L}_{\rm GW}^{\rm Triplet}
&=& \log\mathcal{L}_{\rm GW}^{\rm Doublet}
-\frac{1}{6} \sum_{a,b,c,d} \left\langle h_{,a} \mid h_{,bcd} \right\rangle \Delta\theta^{abcd}
\nonumber \\
&-&\frac{1}{12} \sum_{a,b,c,d,e} \left\langle h_{,ab} \mid h_{,cde} \right\rangle \Delta\theta^{abcde}
-\frac{1}{72} \sum_{a,b,c,d,e,f} \left\langle h_{,abc} \mid h_{,def} \right\rangle \Delta\theta^{abcdef} .
\label{eq:gw_triplet_dali_loglikelihood}
\end{eqnarray}

Here we work with the zero-noise approximation of GW data analysis, where the best-fit parameters $\boldsymbol{\hat{\theta}}$ will be just the fiducial parameters used in the simulations.

Therefore, in the GW formulation, the Fisher approximation is built from products of first waveform derivatives, Doublet adds the terms involving second waveform derivatives, and Triplet further incorporates third waveform derivatives. This structure is particularly useful when the Fisher matrix becomes nearly singular or non-invertible, since higher-order derivative terms can provide a more faithful local approximation to non-Gaussian parameter degeneracies than the purely Gaussian Fisher approximation. In this sense, DALI offers an intermediate approach between analytic Fisher forecasts and full sampling of the exact likelihood \cite{dalichina,gwdali,cosmodali}.
The GW implementation of this formalism was developed in the public Python package \texttt{GWDALI} \cite{gwdali}, which computes Fisher, Doublet, and Triplet approximations for arbitrary networks of terrestrial GW detectors and may interface with waveform models from \texttt{LALSuite} and stochastic sampling through \texttt{Bilby} \cite{gwdali}. Another software which make use of this formalism is the hybrid Mathematica/Python-based code called \texttt{SymDALI}which make use of symbolic derivatives to compute DALI tensors performing stochastic sampling via \texttt{Bilby}~\cite{daSilvaBarbosa:2026zhz} . Ryan et al.~\cite{cosmodali} adapted the same DALI philosophy to cosmological forecasting, using the difference between Fisher and DALI likelihood contours as a diagnostic for the validity of Fisher forecasts. Other developments have extended DALI to parameter-dependent covariance matrices and to variational posterior reconstructions for expensive likelihoods \cite{sellentin2015,sellentin2023}.

In the next section we simulate our GW catalogs in order to build Hubble-diagrams for cosmological inferences.

% <<<<<<<<<<<<<<<<<<<<<<<<<<<<<<<<<<<<<<<<<<<<<<<<<<<<<<<<<<<<<<<<<<<<<<<<<<<<<<<<<<<<<<<<<<<<<<<<<<<<<<<<<<<<<<<<<<<<<<<<<<<

\section{Building the data}

\hspace{1.0cm}The mock catalogs employed in this analysis were constructed using \texttt{GWDALI} v0.1.4. The software was used to generate the simulated source populations, calculate the detector responses, impose the detection criterion, and evaluate the Fisher and DALI expansions considered in the subsequent analysis. The physical assumptions entering the simulations, including the source-population distributions, waveform model, detector network, and fiducial cosmology, are specified independently below.

We generate \(1000\) statistically independent mock catalogs, each containing \(300\) detected GW events with electromagnetic counterparts with network signal-to-noise ratio satisfying $ \rho_{\mathrm{net}}>8.$ Because an electromagnetic counterpart is assumed to be identified for every selected event, the source redshift is treated as directly measured. These catalogs therefore represent bright standard sirens and provide a controlled setting in which the injected source parameters are known. This allows the accuracy and limitations of the Fisher and DALI likelihood approximations to be assessed directly.

\subsection{Source redshift distribution.}

\hspace{0,5cm}The sources are distributed over the redshift interval
\begin{equation}
    z\in[z_{\min},z_{\max}]=[0.02,1].
    \label{eq:redshift_interval}
\end{equation}

The adopted distribution is intended to represent the observable population of bright sirens expected from 3G detectors operating jointly with a dedicated electromagnetic mission such as THESEUS. Considering the parametrization adopted in ~\cite{josiel2022,Belgacem2019}, the redshifts of the sources are sampled from the following normalized distribution
\begin{equation}
    P(z)
    =
    \frac{1}{N_z}
    \frac{z^{3}}
    {1+\exp\!\left(10.6z^{0.6}\right)},
    \qquad
    z_{\min}\leq z\leq z_{\max},
    \label{eq:redshift_distribution}
\end{equation}
where $N_z$ is the normalization constant.

\subsection{Binary population and angular distributions.}

\hspace{0,5cm} The simulated sources are equal-mass compact binaries satisfying
\begin{equation}
    m_{1}=m_{2}= 1.5M_{\odot}.
    \label{eq:equal_mass_assumption}
\end{equation}

The detector-frame masses are obtained from the corresponding source-frame
masses through
\begin{equation}
    m_{i,z}=(1+z)m_{i},
    \qquad i\in\{1,2\}.
    \label{eq:redshifted_component_masses}
\end{equation}

The sky coordinates of the sources ($\alpha$ and $\delta$) are isotropically sampled over the celestial sphere. The polarization angle is sampled uniformly over its appropriate angular interval. These choices ensure that the simulated population does not contain a preferred sky direction.

To approximate the selection effects associated with the observation of an electromagnetic counterpart, the binary inclination angle, $\iota$, is drawn from a smooth-cutoff distribution. Following~\cite{josiel2023}, we adopt
\begin{equation}
P_{\mathrm{smooth}}(\iota)=
\begin{cases}
\sin\iota,
&
\iota\leq\iota_{0}
\ \text{or}\
\iota\geq\pi-\iota_{0},
\\[0.5em]
C_{L}
\exp\!\left[
-\dfrac{1}{2\sigma_{0}^{2}}
\left(
\iota-\iota_{0}
-\dfrac{\sigma_{0}^{2}}{\tan\iota_{0}}
\right)^{2}
\right],
&
\iota_{0}<\iota\leq\dfrac{\pi}{2},
\\[1.0em]
C_{R}
\exp\!\left[
-\dfrac{1}{2\sigma_{0}^{2}}
\left(
\iota-\pi+\iota_{0}
-\dfrac{\sigma_{0}^{2}}
{\tan(\pi-\iota_{0})}
\right)^{2}
\right],
&
\dfrac{\pi}{2}<\iota<\pi-\iota_{0},
\end{cases}
\label{eq:smooth_inclination_distribution}
\end{equation}
where
\begin{align}
    C_{L}
    &=
    \sin\iota_{0}
    \exp\!\left[
    \frac{\sigma_{0}^{2}}
    {2\tan^{2}\iota_{0}}
    \right],
    \label{eq:inclination_normalization_left}
    \\
    C_{R}
    &=
    \sin(\pi-\iota_{0})
    \exp\!\left[
    \frac{\sigma_{0}^{2}}
    {2\tan^{2}(\pi-\iota_{0})}
    \right].
    \label{eq:inclination_normalization_right}
\end{align}
Throughout this work, we adopt
\begin{equation}
    \sigma_{0}=0.1,
    \qquad
    \iota_{0}=20^{\circ}.
    \label{eq:inclination_parameters}
\end{equation}

The parameter \(\iota_{0}\) acts as an effective angular scale for the electromagnetic selection of bright sirens. Joint GW--GRB observations preferentially select binaries viewed close to the orbital angular-momentum axis, corresponding to nearly face-on configurations \((\iota\simeq0)\) or nearly face-off configurations \((\iota\simeq\pi)\) \cite{Arun_2014,Mancarella2024}. In this regime, the GRB viewing angle is closely related to the binary inclination, and the identification of a GRB counterpart therefore favors small values of \(\iota\) or \(\pi-\iota\).

This selection is particularly relevant because the luminosity distance inferred from the GW amplitude is strongly correlated with the inclination angle \cite{Nissanke_2010,Mancarella2024}. Restricting the population to nearly face-on or face-off systems partially reduces this degeneracy and improves the expected luminosity-distance measurements.

The value \(\iota_{0}=20^{\circ}\) is consistent with previous standard-siren forecasts in which angular scales of a few tens of degrees are used to model GRB beaming and electromagnetic selection. Nissanke et al.~\cite{Nissanke_2010}, for example, considered a beamed short-GRB population in which most detectable systems were observed within an angle of approximately \(25^{\circ}\). Similarly, Belgacem et al.~\cite{Belgacem_v2} adopted a GRB beaming angle of approximately \(25^{\circ}\) in mock catalogs of joint GW--GRB standard sirens. The adopted value therefore provides a smooth analytic approximation to the angular preference for bright sirens viewed close to the binary angular-momentum axis, rather than a complete model of electromagnetic detection.\\
\indent The smooth-cutoff inclination distribution adopted here provides an analytic approximation to the angular selection associated with electromagnetic-counterpart detection, avoiding the abrupt transition implied by a hard angular cutoff. However, this prescription does not explicitly model the GRB emission, the instrumental selection function of a specific satellite, or the probability of obtaining a redshift for each event. Instead, we assume that an electromagnetic counterpart and its redshift are available for every source included in the simulated catalogs. A more detailed treatment of these observational effects would be a valuable extension of this work.

\begin{figure}[htbp]
    \centering
    \includegraphics[width=\linewidth]{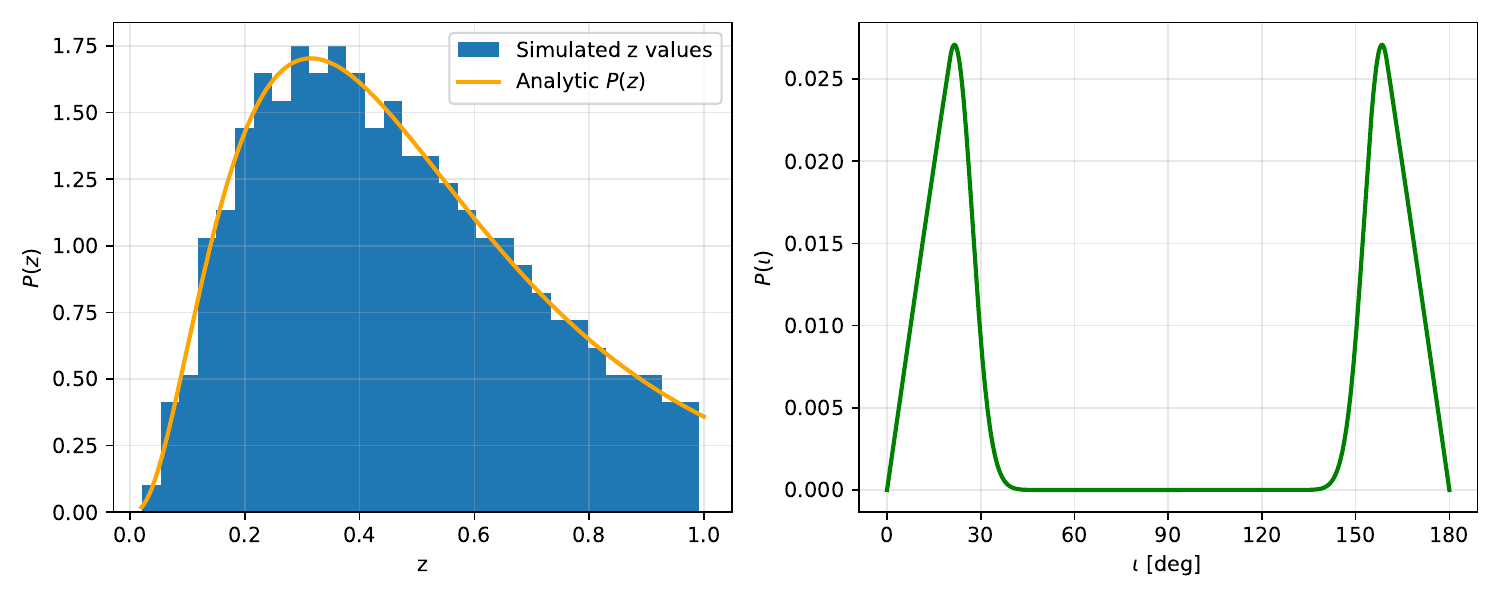}
    \caption{
    Source-population distributions adopted in the mock-catalog simulations.
    Left panel: normalized redshift distribution over
    \(z\in[0.02,1]\). The histogram represents the sampled redshifts, while
    the continuous curve corresponds to the analytic probability density
    Eq.~\eqref{eq:redshift_distribution}. Right panel: smooth-cutoff
    inclination distribution defined in
    Eq.~\eqref{eq:smooth_inclination_distribution}. The distribution
    suppresses nearly edge-on systems and preferentially selects nearly face-on and face-off configurations, approximating the observational   
    selection associated with the detection of a GRB counterpart.
    }
    \label{fig:z_iota_distributions}
\end{figure}

Figure~\ref{fig:z_iota_distributions} summarizes the source-population model used in the simulations. The left panel shows the adopted redshift distribution, while the right panel illustrates the inclination selection associated with the assumed electromagnetic-counterpart population.

\subsection{Gravitational Wave signal model}

\hspace{1.0cm}The GW signals are modeled in the frequency domain using the \texttt{TaylorF2} waveform approximant. TaylorF2 is obtained through the stationary-phase approximation and describes the inspiral phase of a quasicircular compact binary. In the restricted post-Newtonian approximation, the amplitude is retained at leading order, while the Fourier phase is included up to \(3.5\) post-Newtonian order \cite{Damour:2000zb}.

The two gravitational-wave polarizations in the Fourier domain are given by
\begin{align}
    \widetilde{h}_{+}(f)    &=    \sqrt{\frac{5}{24}}\,    \frac{\mathcal{M}_{c,z}^{5/6}}    {\pi^{2/3}d_{L}}\,    f^{-7/6}    \left(    \frac{1+\cos^{2}\iota}{2}    \right)    e^{i\Psi(f)},
    \label{eq:taylorf2_plus}
    \\
    \widetilde{h}_{\times}(f)    &=    i\sqrt{\frac{5}{24}}\,    \frac{\mathcal{M}_{c,z}^{5/6}}    {\pi^{2/3}d_{L}}\,    f^{-7/6}    \cos\iota\,     e^{i\Psi(f)}.
    \label{eq:taylorf2_cross}
    \\
    \Psi(f)    &=    2\pi f t_c    -\phi_c    -\frac{\pi}{4}    +\frac{3}{128\eta   v^5}\sum_{k=0}^7\varphi_kv^k    ,
\end{align}
where $v\equiv(\pi M_z f)^{1/3}$ is the characteristic orbital velocity of the binary, $M_z$ denotes the detector-frame total mass, and $t_c$ and $\phi_c$ are the coalescence time and phase, respectively. The post-Newtonian (PN) coefficients $\varphi_k$ can be found in~\cite{Sathyaprakash:2009xs}. Here, $d_L$ is the luminosity distance, $\iota$ is the inclination angle of the binary, and $\mathcal{M}_{c,z}$ is the detector-frame chirp mass, related to the source-frame chirp mass by
\begin{equation}
\mathcal{M}_{c,z}=(1+z)\mathcal{M}_c,
\qquad
\mathcal{M}_c=\eta^{3/5}M,
\end{equation}
where $M$ is the source-frame total mass and $\eta$ is the symmetric mass ratio.

\subsection{Detector network and detection criterion.}

The simulated detector network consists of ET and CE. CE is modeled as a single L-shaped interferometer, with an opening angle of \(90^\circ\) and the sensitivity curve provided in~\cite{Evans2016}. ET is modeled in its triangular configuration, equivalent to three co-located interferometers with opening angles of \(60^\circ\), rotated by \(0^\circ\), \(120^\circ\), and \(-120^\circ\), respectively. For ET, we adopt the historical ET-D sensitivity curve available in~\cite{ET2018ETD}, as implemented in \texttt{GWDALI}\footnote{A quantitative reassessment using updated ET sensitivity curves, including those adopted in the CoBA Science Study~\cite{ET-0304B-22}, is left for future work.}. The network therefore comprises four detector channels in total.
\begin{table}[ht]
\centering
\begin{tabular}{c c c c c c}
\hline
Detector channel & Observatory & Longitude & Latitude & Rotation & Arm opening angle \\
\hline
\texttt{det0} & CE & \(-119^\circ\) & \(46^\circ\) & \(45^\circ\) & \(90^\circ\) \\
\texttt{det1} & ET & \(10^\circ\) & \(43^\circ\) & \(0^\circ\) & \(60^\circ\) \\
\texttt{det2} & ET & \(10^\circ\) & \(43^\circ\) & \(120^\circ\) & \(60^\circ\) \\
\texttt{det3} & ET & \(10^\circ\) & \(43^\circ\) & \(-120^\circ\) & \(60^\circ\) \\
\hline
\end{tabular}
\caption{Configuration of the third-generation detector network used in this work.}
\label{tab:detector_network}
\end{table}

\begin{figure}[h]
    %\centering
    \includegraphics[width=\linewidth]{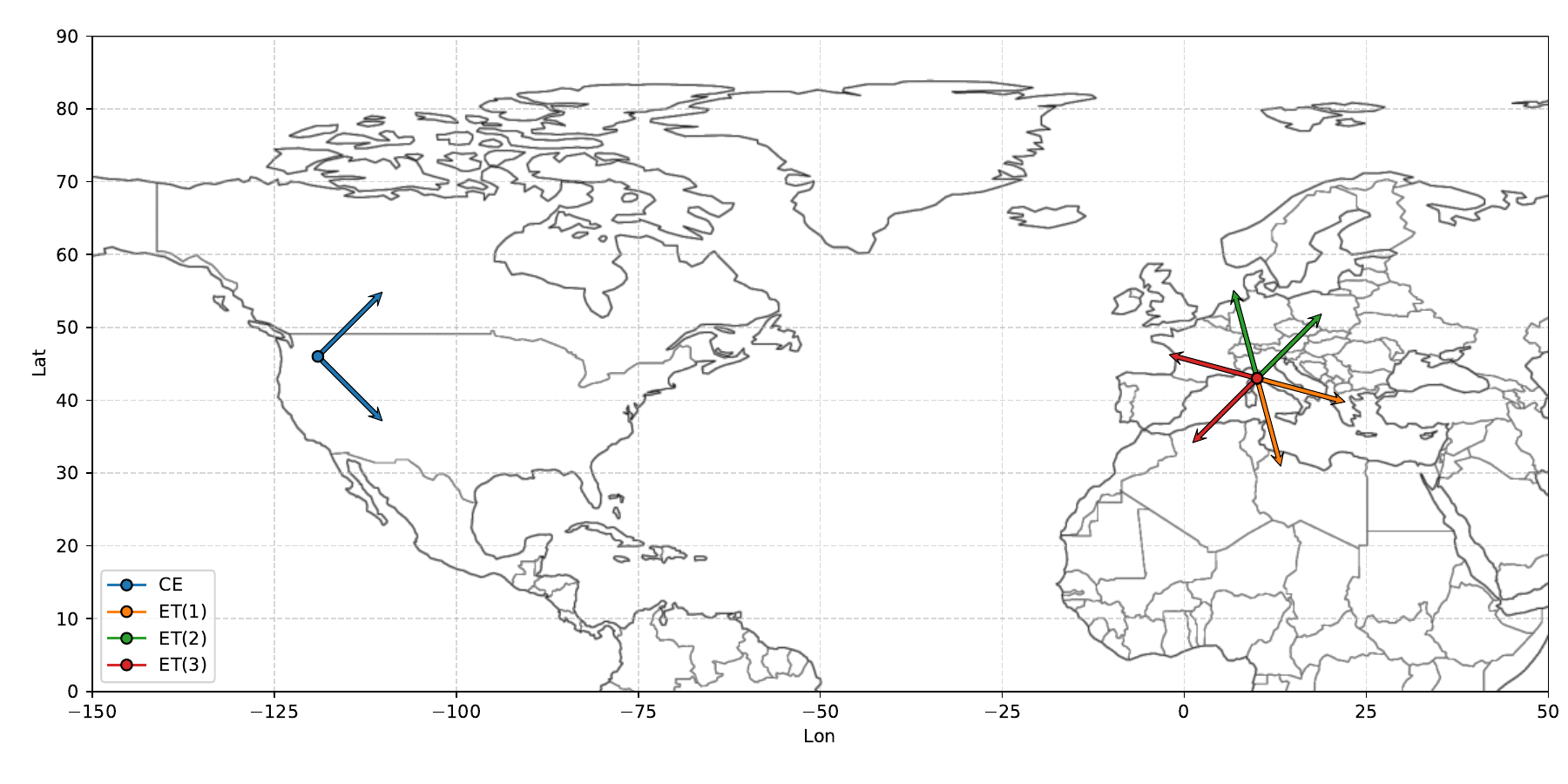}
    \caption{Detector configurations}\label{fig:det_map}
\end{figure}

For detector ``\(a\)", the optimal matched-filter signal-to-noise ratio is
defined as
\begin{equation}
    \rho_{a}^{2} = 4\int_{f_{\min}}^{f_{\max}}
    \frac{\left|\widetilde{h}_{a}(f)\right|^{2}}{S_{n,a}(f)}\,\mathrm{d}f,
    \label{eq:single_detector_snr}
\end{equation}
where \(S_{n,a}(f)\) is the one-sided noise power spectral density of detector \(a\). Here, $\tilde{h}_a(f)$ denotes the GW signal measured at the output of detector $a$, obtained as a linear combination of the two GW polarizations:
\begin{equation}
    \tilde{h}_a(f) =
    \left[
        F_{+,a}\tilde{h}_+(f)
        + F_{\times,a}\tilde{h}_\times(f)
    \right] e^{-2\pi i f \tau_a},
\end{equation}
where $F_{+,a}$ and $F_{\times,a}$ are the antenna-pattern functions, which describe the directional response of detector $a$ to the two GW polarizations. They are given by \cite{Sathyaprakash:2009xs}:
\begin{eqnarray}
    F_{+,a} &=& \sin(\Omega_a)
    \left[
        f_{+,a}\cos(2\psi_a)
        + f_{\times,a}\sin(2\psi_a)
    \right], \\
    F_{\times,a} &=& \sin(\Omega_a)
    \left[
        -f_{+,a}\sin(2\psi_a)
        + f_{\times,a}\cos(2\psi_a)
    \right], \\
    f_{+,a} &=& \frac{1}{2}(1+\cos^2\beta_a)\cos(2\alpha_a), \\
    f_{\times,a} &=& -\cos\beta_a\sin(2\alpha_a).
\end{eqnarray}
Here, $\Omega_a$ denotes the opening angle between the interferometer arms, while $\alpha_a$ and $\beta_a$ specify the source sky position in spherical coordinates defined with respect to detector $a$. The angle $\psi_a$ is the GW polarization angle in the detector's reference frame. Finally, $\tau_a$ denotes the GW arrival-time delay at detector $a$ relative to a chosen reference frame, such as the Earth's center or another detector.

Thus, the network signal-to-noise ratio is
\begin{equation}
    \rho_{\mathrm{net}}
    =
    \left(
    \sum_{a=1}^{N_{\mathrm{det}}}
    \rho_{a}^{2}
    \right)^{1/2}.
    \label{eq:network_snr}
\end{equation}
Candidate sources are generated until each catalog contains \(300\) events
satisfying the detection condition $\rho_{net}>8$.

\subsection{Fiducial cosmology}

\hspace{1.0cm}The luminosity distance associated with each simulated redshift is calculated assuming a spatially flat \(\omega_0\omega_a\)CDM cosmology. The dark-energy equation of state is parametrized as
\begin{equation}
    \omega(z) = \omega_0 + \omega_a\frac{z}{1+z}.
    \label{eq:cpl_equation_of_state}
\end{equation}
For a flat Universe, the dimensionless expansion rate is
\begin{equation}
    E(z)
    =
    \left[
    \Omega_{m}(1+z)^{3}
    +
    \left(1-\Omega_{m}\right)
    (1+z)^{3(1+\omega_0+\omega_a)}
    \exp\!\left(
    -\frac{3\omega_az}{1+z}
    \right)
    \right]^{1/2},
    \label{eq:dimensionless_Hubble_rate}
\end{equation}

The fiducial cosmological parameters used to generate the catalogs are
\begin{equation}
    H_{0} = 70\,\mathrm{km\,s^{-1}\,Mpc^{-1}},
    \qquad \Omega_{m}=0.3, \qquad \Omega_{\Lambda}=0.7, \qquad  \omega_{0}=-1, \qquad \omega_a=0\ .
\end{equation}

% ==================================================
% =================================================
\section{GW Inference}
\label{sec:gw_inference}

\hspace{1.0cm} The GW analysis applies the \texttt{GWDALI} package \cite{gwdali}. Following the specifications introduced in the previous section, the CBC signals are modeled in the frequency domain using the TaylorF2 approximant. The set of free GW parameters entering the analysis is

\begin{equation}
\boldsymbol{\theta}_{\mathrm{GW}}=\left\{\mathcal{M}_{c},\eta,X,\iota,\psi,\phi_{\mathrm{coal}}\right\},
\label{eq:thetagw}
\end{equation}
where $\psi$ is the polarization angle, and $\phi_{\mathrm{coal}}$ is the coalescence phase. The quantity $X$ denotes the luminosity-distance variable and is taken to be either $d_L$ or $1/d_L$, depending on the adopted inference prescription.

Since the Fisher matrix provides a local quadratic approximation to the likelihood, the way in which this information is converted into a luminosity-distance measurement can have a non-negligible impact on the resulting GW catalogue. We therefore consider several Fisher-based prescriptions, differing in the distance parametrization, matrix inversion, inclusion of prior information, and posterior construction.

\subsection{Fisher-Based Approaches}
\label{sec:fisher_approaches}

\hspace{1.0cm} We consider three different prescriptions for the Fisher information. The first corresponds to the original GWDALI implementation used as our baseline, while the remaining approaches were introduced to investigate the dependence of the inferred luminosity-distance uncertainties on the treatment of poorly conditioned Fisher matrices and on the adopted prior information.

\paragraph{Direct Fisher inversion.}
We construct the Fisher matrix using the inverse luminosity distance, $y\equiv 1/d_L$, and subsequently transform it to the $d_L$ parametrization before inversion. Since

\begin{equation}
\frac{\partial y}{\partial d_L}=-\frac{1}{d_L^2} =-y^2,
\label{eq:jacobian_invdl}
\end{equation}
the Fisher elements involving the distance parameter transform as

\begin{equation}
\Gamma_{d_Ld_L}=y^4\Gamma_{yy},\qquad\Gamma_{d_L\theta}=-y^2\Gamma_{y\theta},
\label{eq:fisher_dl_transformation}
\end{equation}
where $\theta$ denotes any of the remaining GW parameters. The resulting matrix is rescaled by its diagonal elements to improve numerical conditioning and is then inverted using the ordinary matrix inverse,

\begin{equation}
\Sigma_{\mathrm{F}}=\Gamma_{\mathrm{F}}^{-1}.
\label{eq:fisher_direct_inverse}
\end{equation}

The reliability of each inversion is explicitly assessed through

\begin{equation}
\epsilon_{\mathrm{inv}}=\max_{ij}\left|\left(\Sigma_{\mathrm{F}}\Gamma_{\mathrm{F}}-I\right)_{ij}\right|,
\label{eq:fisher_inverse_error}
\end{equation}
and only events satisfying $\epsilon_{\mathrm{inv}}<0.05$ are retained. Singular matrices, matrices with invalid diagonal elements, and covariance matrices with non-finite or negative diagonal elements are also rejected.

\paragraph{Fisher matrix with Gaussian priors.}
In this prescription, the same transformation to the $d_L$ parametrization is first applied. Independent Gaussian priors are then introduced to regularize poorly constrained directions in parameter space. Their contribution to the Fisher matrix is written as

\begin{equation}
\Gamma_{\mathrm{prior}}=\operatorname{diag}\left(\frac{1}{\sigma_{d_L}^{2}},
\frac{1}{\sigma_{\iota}^{2}},
\frac{1}{\sigma_{\psi}^{2}},
\frac{1}{\sigma_{\phi_{\mathrm{coal}}}^{2}},
\frac{1}{\sigma_{\mathcal{M}_{c}}^{2}},
\frac{1}{\sigma_{\eta}^{2}}
\right),
\label{eq:fisher_prior}
\end{equation}
with prior widths

\begin{equation}
\boldsymbol{\sigma}_{\mathrm{prior}}=\left(100,\pi,\pi,\pi,100,0.25\right),
\label{eq:prior_widths}
\end{equation}
following the parameter ordering
$\{d_L,\iota,\psi,\phi_{\mathrm{coal}},\mathcal{M}_c,\eta\}$.
The total information matrix is therefore

\begin{equation}
\Gamma_{\mathrm{tot}}=\Gamma_{\mathrm{F}}+\Gamma_{\mathrm{prior}}.
\label{eq:fisher_total_prior}
\end{equation}

The covariance matrix is obtained using the same diagonal rescaling and direct-inversion procedure adopted above,

\begin{equation}
\Sigma=\Gamma_{\mathrm{tot}}^{-1},
\label{eq:fisher_prior_covariance}
\end{equation}
and the same criterion $\epsilon_{\mathrm{inv}}<0.05$ is applied to assess the numerical reliability of the inversion.

For the accepted events, the Fisher posterior is represented by a multivariate Gaussian centered on the injected source parameters,

\begin{equation}
p\left(\boldsymbol{\theta}_{\mathrm{GW}}\right)\simeq\mathcal{N}\left(\boldsymbol{\theta}_{\mathrm{true}},\Sigma
\right),\label{eq:fisher_gaussian_posterior}
\end{equation}
from which $10^{4}$ samples are directly generated. No MCMC sampling is employed in this procedure.

\paragraph{Singlet-DALI.}
Following \cite{josiel2022}, a conceptually different prescription is adopted in which the Fisher matrix is used to define a Gaussian approximation to the event likelihood,

\begin{equation}
\ln\mathcal{L}_{\mathrm{F}}(\boldsymbol{\theta})=\ln\mathcal{L}_{0}-\frac{1}{2}\Delta\boldsymbol{\theta}^{T}\Gamma_{\mathrm{F}}
\Delta\boldsymbol{\theta},
\label{eq:fisher_likelihood}
\end{equation}
where
$\Delta\boldsymbol{\theta}=\boldsymbol{\theta}-\boldsymbol{\theta}_{0}$
and $\boldsymbol{\theta}_{0}$ denotes the injected parameters. The corresponding posterior is then obtained through

\begin{equation}
p(\boldsymbol{\theta}\mid d)\propto\mathcal{L}_{\mathrm{F}}(\boldsymbol{\theta})\pi(\boldsymbol{\theta}),
\label{eq:singlet_posterior}
\end{equation}
and explored numerically using Monte Carlo sampling. In contrast to the previous approaches, the posterior samples are obtained by sampling the Fisher likelihood together with the adopted priors, rather than being drawn directly from the covariance matrix. This procedure is sometimes referred to as \emph{Fisher sampling}, \emph{MCMC Fisher}, or \emph{Singlet-DALI} \cite{josiel_2026}.

\subsection{Doublet-DALI}
\label{sec:doublet_dali}

\begin{figure}
     \centering
     \includegraphics[width=1.0\linewidth]{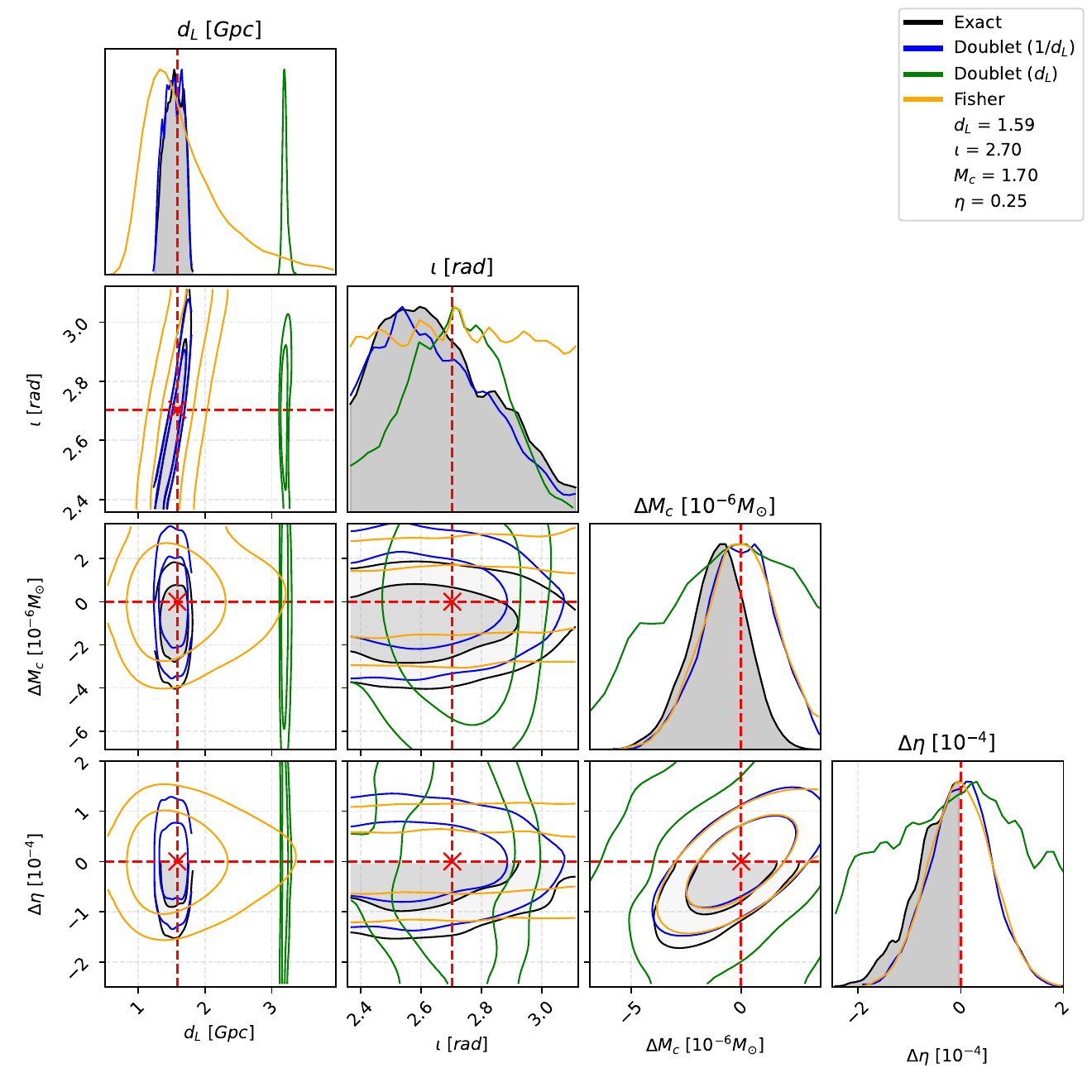}
     \caption{Joint constraints on GW parameters for several likelihood approximations. For better visualizations we denote $\Delta M_c\equiv M_c-M_c^{true}$ and $\Delta \eta \equiv \eta-\eta^{true}$ as the displacements from the injection parameters. The Doublet approximation with prior $p(d_L)$ exhibits a secondary mode at larger $d_L$, reflecting the degeneracy between luminosity distance and inclination. In contrast, using the inverse-distance prior $p(1/d_L)$ re-weights the distance distribution and suppresses this secondary mode, yielding constraints more concentrated around the main posterior region.}
     \label{fig:dL_GWDALI}
 \end{figure}

\hspace{1.0cm} Beyond the Fisher approximation, we also consider the Doublet-DALI expansion. For the Doublet calculation, we adopt $1/d_L$ as the distance variable. This reparametrization is motivated by the behavior of the approximate likelihood in the $(d_L,\iota)$ sector: when the expansion is performed directly in $d_L$, the Doublet reconstruction can exhibit a pronounced bimodality associated with the luminosity-distance--inclination degeneracy. Expressing the likelihood in terms of $u=d_L^{-1}$ leads to a smoother local structure and improves the numerical stability of the approximation \cite{josiel_2026}, as illustrated in Figure \ref{fig:dL_GWDALI}. 

A summary of the priors adopted in the Doublet GW analyses is presented in Table~\ref{tab:gwdali_priors}. The priors for \(\mathcal{M}_c\), \(\eta\), \(\psi\), \(\phi_{\rm coal}\), and \(\iota\) are common to both analyses, while the distance-related prior is implemented directly in \(d_L\) for the Fisher analysis and in \(u\equiv d_L^{-1}\) for the Doublet analysis. For the inclination angle, we adopt a conservative truncated prior, motivated by the fact that the population of bright sirens is expected to be affected by the selection of events with detectable electromagnetic counterparts and therefore need not follow the isotropic inclination distribution of the underlying binary population. Rather than adopting the inclination distribution of the simulated injections as the prior, we use the same broad, truncated prior for all events. This choice avoids encoding the specific inclination distribution assumed in the simulations directly into the inference and reflects the fact that the inclination distribution of the observed bright-siren population is not known a priori.

\begin{table}[htbp]
\centering
\renewcommand{\arraystretch}{1.25}
\begin{tabular}{p{0.16\textwidth} p{0.40\textwidth} p{0.22\textwidth}}  %p{0.34\textwidth}}
\hline
\textbf{Parameter} & \textbf{Prior range} & \textbf{Prior type} \\ %\textbf{Use / implementation} \\
\hline

\(\mathcal{M}_c\) &
\(\mathcal{M}_{c,\ast}(1-10^{-3}) \leq \mathcal{M}_c \leq
\mathcal{M}_{c,\ast}(1+10^{-3})\) &
Uniform \\
%The prior is centered on the fiducial chirp mass
%of each event and sampled with 100 grid points. \\

\(\eta\) &
\(\eta_{\ast}(1-10^{-3}) \leq \eta \leq
\eta_{\ast}(1+10^{-3})\) &
Uniform  \\
%The prior is centered on the fiducial symmetric
%mass ratio of each event and sampled with 100 grid points. \\

\(\psi\) &
\([0,\pi]\) &
Uniform  \\
 %with \(p(\psi)\propto 1\). \\

\(\phi_{\rm coal}\) &
\([-\pi,\pi]\) &
Uniform  \\
 %with \(p(\phi_{\rm coal})\propto 1\). \\

\(\iota\) &
\([0,\pi]\) &
Sine with cutoff between [$45^o,\ 135^o$]  \\
 %with \(p(\iota)\propto 1\). \\

\(d_L\) &
%Constructed from \(z\in[0.01,5.0]\) &
[$4\cdot 10^{-3},\ 46.7$] Gpc $\sim z\in[0.01,5.0]$ &
Comoving-volume prior \\
%Used in the Fisher analysis. The prior is built from the luminosity distance
%\(d_L(z)\), the Hubble parameter \(H(z)\), and the comoving distance
%\(R_c(z)=d_L(z)/(1+z)\). The unnormalized density is taken as
%\(p_{\rm raw}(z)\propto R_c^2(z)/H(z)\), then reordered as a function of
%\(d_L\) and normalized in \(d_L\). \\

\(u\equiv d_L^{-1}\) &
\(u\in[1/d_{L,\max},1/d_{L,\min}]\) &
Transformed comoving-volume prior \\ \hline
%Used in the Doublet analysis. The comoving-volume prior is first constructed in
%\(d_L\), then mapped to \(u=d_L^{-1}\). The Jacobian
%\(\left|dd_L/du\right|=d_L^2\) is applied, and the resulting prior is normalized
%in \(u\). \\
\end{tabular}
\caption{Priors adopted in the Doublet \texttt{GWDALI} analysis.}
\label{tab:gwdali_priors}
\end{table}

Posterior samples are then obtained with a nested-sampling algorithm and are used to characterize the uncertainty associated with the distance-related parameter inferred from each event. The main numerical inputs used in \texttt{GWDALI} are listed in Table~\ref{tab:gwdali_inputs}.

\begin{table}[ht]
\centering
\begin{tabular}{p{0.32\textwidth} p{0.48\textwidth}}
\hline
\textbf{Input} & \textbf{Value used} \\
\hline
\texttt{approximant} & \texttt{TaylorF2\_py} \\
\texttt{sampler\_method} & \texttt{nestle} \\
\texttt{npoints} & \(300\) \\
\texttt{step\_size} & \(10^{-5}\) \\
\texttt{rcond} & \jm{0} \\
\texttt{detectors} & \texttt{det0, det1, det2, det3} \\
\hline
\end{tabular}
\caption{Main numerical inputs used in the \texttt{GWDALI} run.}
\label{tab:gwdali_inputs}
\end{table}

\subsection{Hubble Diagram from GWs via Fisher/DALI}

\subsubsection{Fisher Enhancements for GW}
\label{sec:fisher_enhancements}

\begin{figure}[h]
        \centering
        \includegraphics[width=\linewidth]{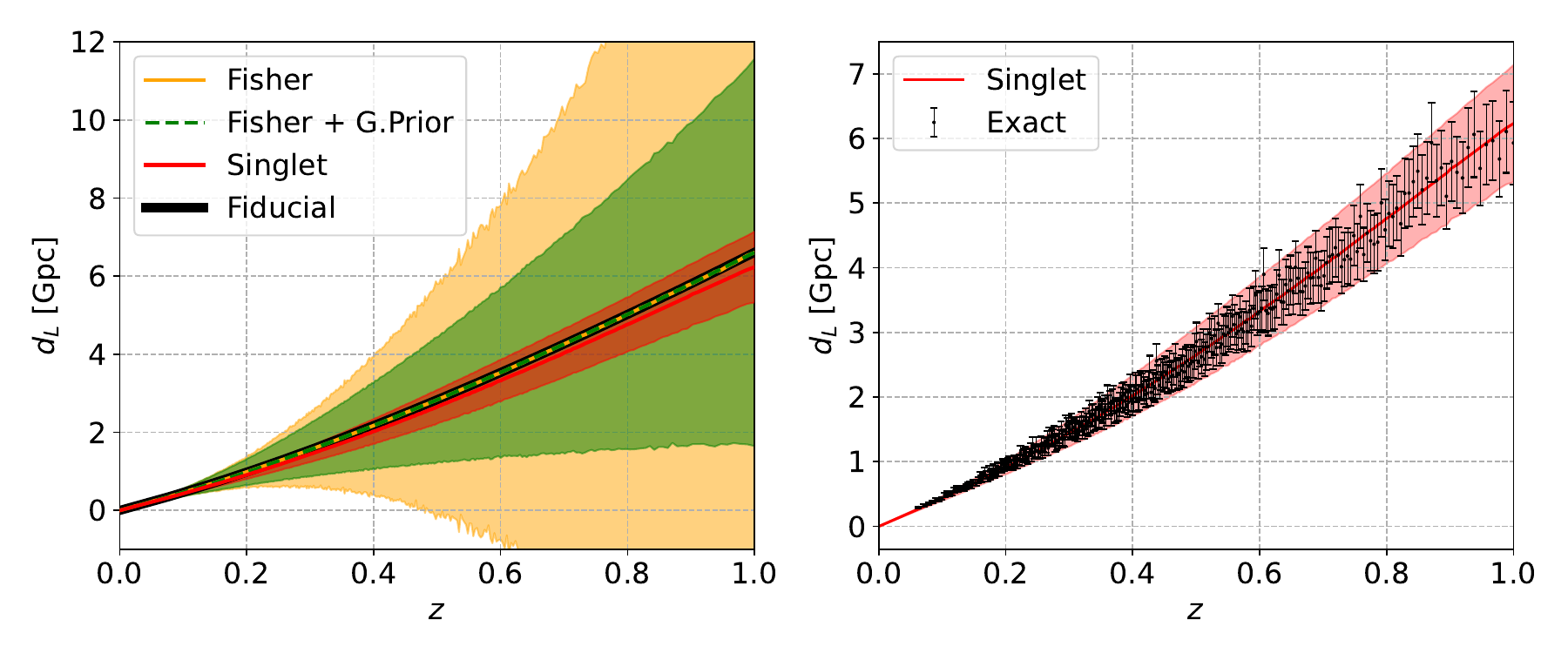}
        \caption {Measurements and uncertainties averaged over 1000 catalogs. Comparisons among different Fisher treatments (left). Comparison between averaged Singlet measurement and a single exact catalog (right).}
        \label{fig:FisherApproachs}
    \end{figure}

\hspace{1.0cm} The different Fisher-based prescriptions described in Sec.~\ref{sec:fisher_approaches} were first compared at the level of the reconstructed Hubble diagram. Figure~\ref{fig:FisherApproachs} shows the luminosity-distance measurements obtained with the direct Fisher inversion, the Fisher matrix with Gaussian priors, and the Singlet approach. The measurements and uncertainties shown in the left panel are averaged over the 1000 simulated catalogs, allowing the systematic behavior of each prescription to be compared over the full redshift interval.

The direct inversion of the Fisher matrix produces the largest luminosity-distance uncertainties among the considered prescriptions. Although only events satisfying the inversion criterion $\epsilon_{\rm inv}<0.05$ are retained, the marginalized uncertainties increase rapidly with redshift. This behavior follows from the poorly constrained directions of the GW parameter space that remain present after the rejection of numerically unreliable inversions, particularly those associated with the luminosity-distance--inclination correlation.

Adding Gaussian prior information to the Fisher matrix modifies the treatment of weakly constrained parameter directions before inversion. The resulting uncertainties are smaller than those obtained with the direct Fisher inversion. The Gaussian priors therefore regularize the covariance estimation without reproducing the golden-event population observed in the pseudo-inverse prescription.

The Singlet approach differs from the previous prescriptions because the Fisher matrix is not inverted and interpreted directly as the final posterior covariance. Instead, the quadratic Fisher likelihood is sampled together with the adopted physical priors. Among the Fisher-based treatments considered here, this prescription produces the luminosity-distance reconstruction closest to that obtained from the exact GW likelihood. The right panel of Fig.~\ref{fig:FisherApproachs} compares the Hubble diagram obtained from the ensemble-averaged Singlet catalogs with the exact reconstruction from a representative catalog, showing consistent distance measurements over the analyzed redshift range.

These results demonstrate that the construction of a Fisher-based Hubble diagram depends not only on the quadratic approximation to the GW likelihood, but also on how this approximation is converted into marginalized parameter constraints. All approaches lead to distinct luminosity-distance uncertainties even when they originate from the same local Fisher information. Among these treatments, Fisher sampling provides the closest agreement with the exact GW reconstruction and is consequently adopted as the reference Fisher enhancement in the subsequent comparisons.

\begin{figure}[h]
    \centering
    \includegraphics[width=\linewidth]{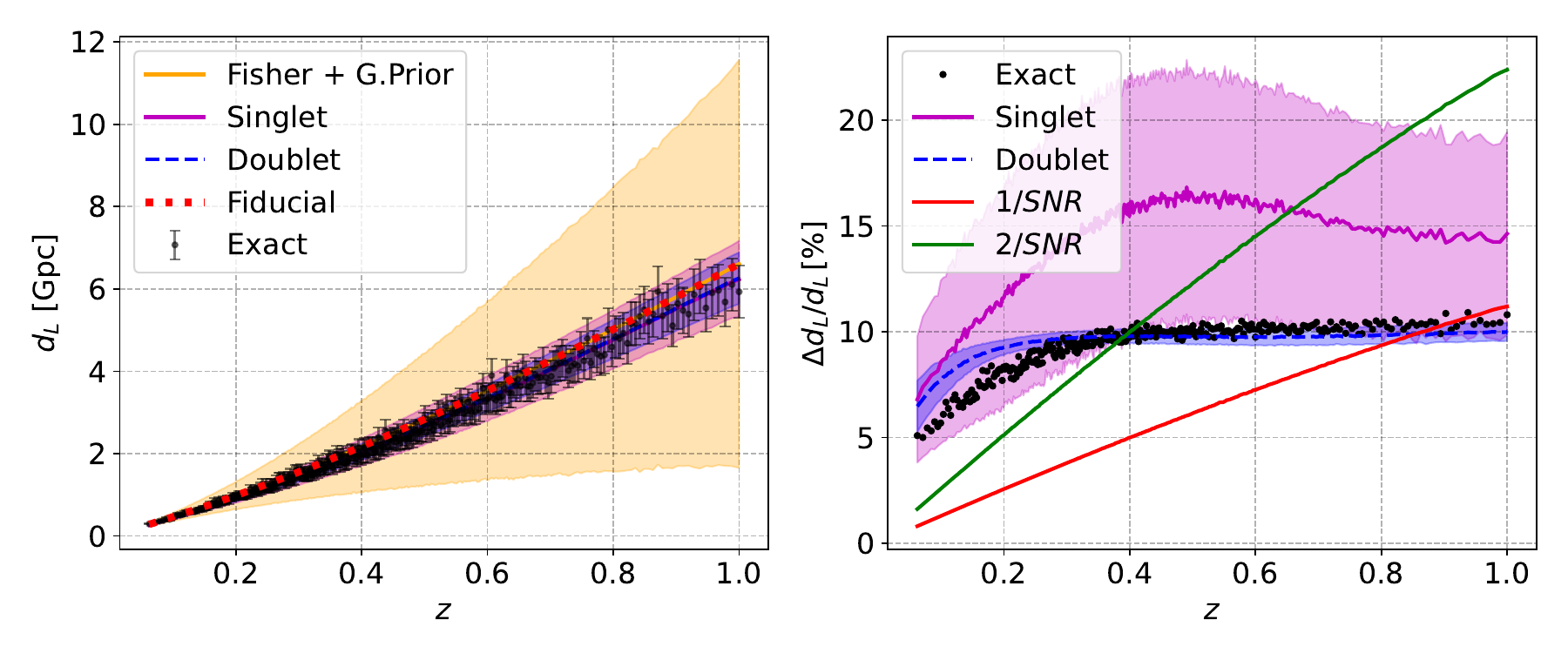}
    \caption{Hubble diagram including average over GW-Singlet and GW-Doublet catalogs.}
    \label{fig:comparison_doublet}
\end{figure}

The comparison with a single GW-Exact realization shown in Fig.~\ref{fig:FisherApproachs} is intended only as an illustrative reference, providing a visual indication of how the mean reconstructed quantities from the approximate catalogs compare with an individual exact realization. A systematic validation of the GW Fisher and DALI approximations against the exact GW likelihood was presented in our previous work on the validation and release of GWDALI v1.0~\cite{josiel_2026}. Here, we focus instead on the propagation of the GW approximations to cosmological inference across the simulated catalog ensembles.

%%%%%%%%%%%%%%%%%%%%%%%%%%%%%%%%%%%%%%%%%%%%%%%%%%%%%%%%%%%%%%%%%%%%%%%%
%%%%%%%%%%%%%%%%%%%%%%%%%%%%%%%%%%%%%%%%%%%%%%%%%%%%%%%%%%%%%%%5%%%%%%%%
% \subsubsection{Fisher Enhancements for GW}
%
% 100, np.pi, np.pi, np.pi, 100, 0.25
%
%     \jm{ See Fig.~\ref{fig:FisherApproachs}
%     \begin{itemize}
%         \item Pure Fisher inversion accepting only inversions with conditioning number $\epsilon\equiv max\, |F\cdot F^{-1}-\mathbb{I}| <0.05$ (\textbf{Results:} huge uncertainties in $d_L$);
%         \item Pseudo-inversion of More-Penrose with $e_{vals}<r_{cond}\equiv10^{-4}$ (\textbf{Results:} Bad $\epsilon$ and generation of \textit{golden-events} - sources with tiny uncertainties in $d_L$);
%         \item Inclusion of Gaussian priors above Fisher $F_G\equiv F +F_{\pi}$ with, $$F_\pi^{-1}=diag(\Delta d_L^2,\ \Delta\iota^2,\ \Delta\psi^2,\ \Delta\phi_0^2,\ \Delta \mathcal{M}_c^2,\ \Delta\eta^2)\ ,$$\footnote{\jm{Here I used $\Delta d_L= 100\, Gpc$, $\Delta\iota=\Delta\psi=\Delta\phi_{coal}=\pi$, $\Delta \mathcal{M}_c=100\, M_{\odot}$ and $\Delta\eta=0.25$.}} 
%         results in good conditioning numbers $\epsilon$ (\textbf{Results:} Smaller uncertainties in $d_L$ in comparison with pure inversions but generating no golden-events);
%         \item Fisher sampling approach via MCMC (we call it \textbf{Singlet} in the spirit of DALI approach) using arbitrary priors (\textbf{Results:} Much better results approaching the exact results - the best Fisher enhancement).
%     \end{itemize}
%     }

\subsubsection{Doublet Approach}

\hspace{1.0cm}The reconstructed mean distance--redshift relation shown in Fig.~\ref{fig:comparison_doublet} reveals a qualitatively different behavior between the Fisher (different approaches) and Doublet approximations. The Fisher reconstruction remains approximately centered on the fiducial relation across the full redshift range, although its uncertainty band broadens rapidly with increasing redshift. In contrast, the Doublet reconstruction produces substantially narrower uncertainty bands, but its mean is systematically displaced toward smaller luminosity distances.

This comparison therefore highlights a trade-off between precision and accuracy in the present implementation. While the Doublet approximation better constrains the local shape of the likelihood, the resulting reduction in posterior width is accompanied by a bias in the recovered mean distance. Since the DALI tensors are computed through automatic differentiation, this behavior is more plausibly associated with the truncation of the expansion at the Doublet level, for which a local higher-order approximation may still be insufficient to reproduce the global structure of a strongly non-Gaussian posterior.

The discrepancy becomes more pronounced at higher redshift, where the inference progressively departs from the high signal-to-noise regime in which the Fisher approximation is expected to perform best \cite{dalichina}. In this regime, the $d_L$--$\iota$ degeneracy becomes increasingly important, making the inferred luminosity distance more sensitive to differences in the local likelihood approximation. The recovery of $d_L$ also depends on the detector geometry, the network capability to resolve the GW polarizations, and the adopted prior on $\iota$ \cite{josiel2023}. The downward displacement observed for the Doublet reconstruction can therefore be interpreted as a consequence of applying a truncated local expansion to an inference problem with significant non-linear parameter correlations.

Taken together, these Figs. ~\ref{fig:FisherApproachs}, ~\ref{fig:comparison_doublet} indicate complementary limitations of the two approximations. The Fisher matrix fails to adequately capture the joint $(d_L,\iota)$ structure in the present scenario, as its purely quadratic form leads to artificial angular singularities and increasingly large uncertainties at high redshift. The Doublet approximation, in turn, provides a more regular description of the likelihood geometry and significantly reduces the inferred uncertainties, but introduces a redshift-dependent bias in the recovered luminosity distance. In this context, DALI can be regarded as a controlled extension of the Fisher formalism, capable of capturing relevant non-Gaussian corrections and elucidating the limitations of a purely ellipsoidal approximation to the likelihood \cite{sellentin2014,cosmodali}.

\section{Cosmological Constraints}
\subsection{Cosmological Model}

\hspace{1.0cm} The cosmological analysis treats the simulated GW detections as bright standard sirens. The redshift of each event is assumed to be known exactly from the electromagnetic counterpart and is taken directly from the mock catalogs. The luminosity-distance information and its associated uncertainty are inherited from the GW analysis described above.
The cosmological parameter vector is defined as
\begin{equation}
\theta_{\mathrm{cosmo}}=\{H_{0},\Omega_{m},\omega_{0},\omega_{a}\},
\label{eq:thetacosmo}
\end{equation}
corresponding to the CPL parametrization of the dark-energy equation of state ($\omega_{DE}(a) = \omega_0 + (1-a)\omega_a$ where $a$ is the scale factor). The Hubble expansion rate is written as
\begin{equation}
H(z)=H_{0}
\sqrt{
\Omega_{m}(1+z)^{3}
+
(1-\Omega_{m})(1+z)^{3(1+\omega_{0}+\omega_{a})}
\exp\!\left(-\frac{3\omega_{a}z}{1+z}\right)
},
\label{eq:hz}
\end{equation}
and the luminosity distance is given by
\begin{equation}
d_{L}(z;\theta_{\mathrm{cosmo}})
=
(1+z)\int_{0}^{z}\frac{dz'}{H(z';\theta_{\mathrm{cosmo}})}.
\label{eq:dlcosmo}
\end{equation}

For a catalog of GW sources, the cosmological log-likelihood is taken to be
\begin{equation}
\log \mathcal{L}_{\mathrm{cosmo}}(\theta_{\mathrm{cosmo}})
=
-\frac{1}{2}
\sum_{n}
\frac{
\left[d_{L,n}^{\mathrm{obs}}-d_{L}(z_{n};\theta_{\mathrm{cosmo}})\right]^{2}
}{
\sigma_{d_L,n}^{2}
},
\label{eq:loglcosmo}
\end{equation}
where $d_{L,n}^{\mathrm{obs}}$ and $\sigma_{d_L,n}$ denote, respectively, the luminosity distance inferred from the GW reconstruction and its associated uncertainty.

\subsection{Semi-analytic Inference}
\hspace{1.0cm}To understand how the measurements of $d_L$ directly affect the measurement of $H_0$, we approximate the luminosity-distance relation by its first-order cosmographic expansion in redshift, as follows:
\begin{equation}
    d_L(z) \approx \frac{z}{H_0}\, .
\end{equation}

Let us also to approximate our likelihood via Fisher approximation. In this case, we have a single 1D Fisher Matrix given by:
\begin{equation}
    F = -\frac{\partial^2 log \mathcal{L}}{\partial H_0^2} = \frac{1}{H_0^2}\sum_i\left(\frac{d_{L,i}}{\sigma_{d_L,i}}\right)^2\, ,
\end{equation}
leading to the uncertainty in $H_0$ as:
\begin{equation}
    \sigma_{H_0} = F^{-1/2} = \frac{H_0}{\sqrt{\sum_i (d_{L,i}/\sigma_{d_L,i})^2}}\, .
\end{equation}

Hence, when we have a single GW event, the uncertainty in $H_0$ has a beautiful relation with respect to $d_L$ given by:
\begin{equation}
    \frac{\sigma_{H_0}}{H_0} = \frac{\sigma_{d_L}}{d_L} \, .
\end{equation}

Such a result implies tell us that the better the distance measurements, the better the $H_0$ measurements.

\subsection{Numerical Inference}

\begin{table}[h]
\centering
%\scriptsize
\setlength{\tabcolsep}{6pt}
\renewcommand{\arraystretch}{1.08}
\begin{tabular}{c c c}
\hline
\textbf{Parameter} & \textbf{Prior} & \textbf{Range} \\
\hline
\(H_0\) & Uniform & \([50,90]\) \\
\(\Omega_m\) & Uniform & \([0.01,0.9]\) \\
\(\omega _0\) & Uniform & \([-5,5]\) \\
\(\omega _a\) & Uniform & \([-10,10]\) \\
\hline
\end{tabular}
\caption{Uniform priors adopted in the \texttt{Bilby} cosmological analysis.}
\label{tab:bilby_priors}
\end{table}

For the cosmology case, the likelihood is approximated through derivative expansions around the best-fit, calculated. The derivatives of the luminosity-distance function with respect to the cosmological parameters are evaluated using automatic differentiation (using the \texttt{Python} package \texttt{JAX}~\cite{jax2018github}), rather than finite-difference schemes, ensuring machine-precision derivatives and avoiding the need to choose numerical step sizes.

The approximate cosmological likelihood is sampled with \texttt{Bilby}, adopting uniform priors on all cosmological parameters. The priors are taken to be broad enough, so that they do not artificially truncate the posterior. The posterior samples were obtained with the \texttt{nestle} nested-sampling algorithm implemented in \texttt{Bilby}, using \texttt{npoints}=1000. The uniform prior ranges adopted for the cosmological parameters are listed in Table~\ref{tab:bilby_priors}, where \(H_0\) is given in \({\rm km\,s^{-1}\,Mpc^{-1}}\).

The cosmological inference is performed cumulatively as a function of redshift. For this purpose, the total interval \(z\in[0,1]\) is divided into ten cumulative redshift intervals, with upper limits \(z_{\max}=0.1,0.2,\ldots,1.0\). For each maximum redshift \(z_{\max}\), we select all sources with \(z_i<z_{\max}\), so that the number of events increases as the redshift upper limit is raised. The cumulative catalogs used at a given redshift threshold is therefore defined as $\mathcal{D}(z_{\max})=\left\{d_{L,i}^{\rm obs},\sigma_{d_L,i},z_i\; \big| \; z_i<z_{\max}\right\}$. Thus, the inference at \(z_{\max}=0.1\) uses only the nearest sources, while higher redshift thresholds include all previously selected events plus the additional sources located between the previous and current redshift cuts.

In addition to the Fisher and Doublet approximations, we also perform the cosmological inference using the exact likelihood for the luminosity distance data, following equation ~\eqref{eq:loglcosmo}. The Fisher and Doublet likelihoods are instead obtained from the corresponding local derivative expansion around the peak of marginalized posteriors. For each value of \(z_{\max}\), the exact, Fisher, and Doublet likelihoods are sampled independently with \texttt{Bilby} using ``nestle" with \texttt{nlive=300}.

\label{sec:results}

\begin{figure}[h]
    \centering
    \includegraphics[width=0.7\linewidth]{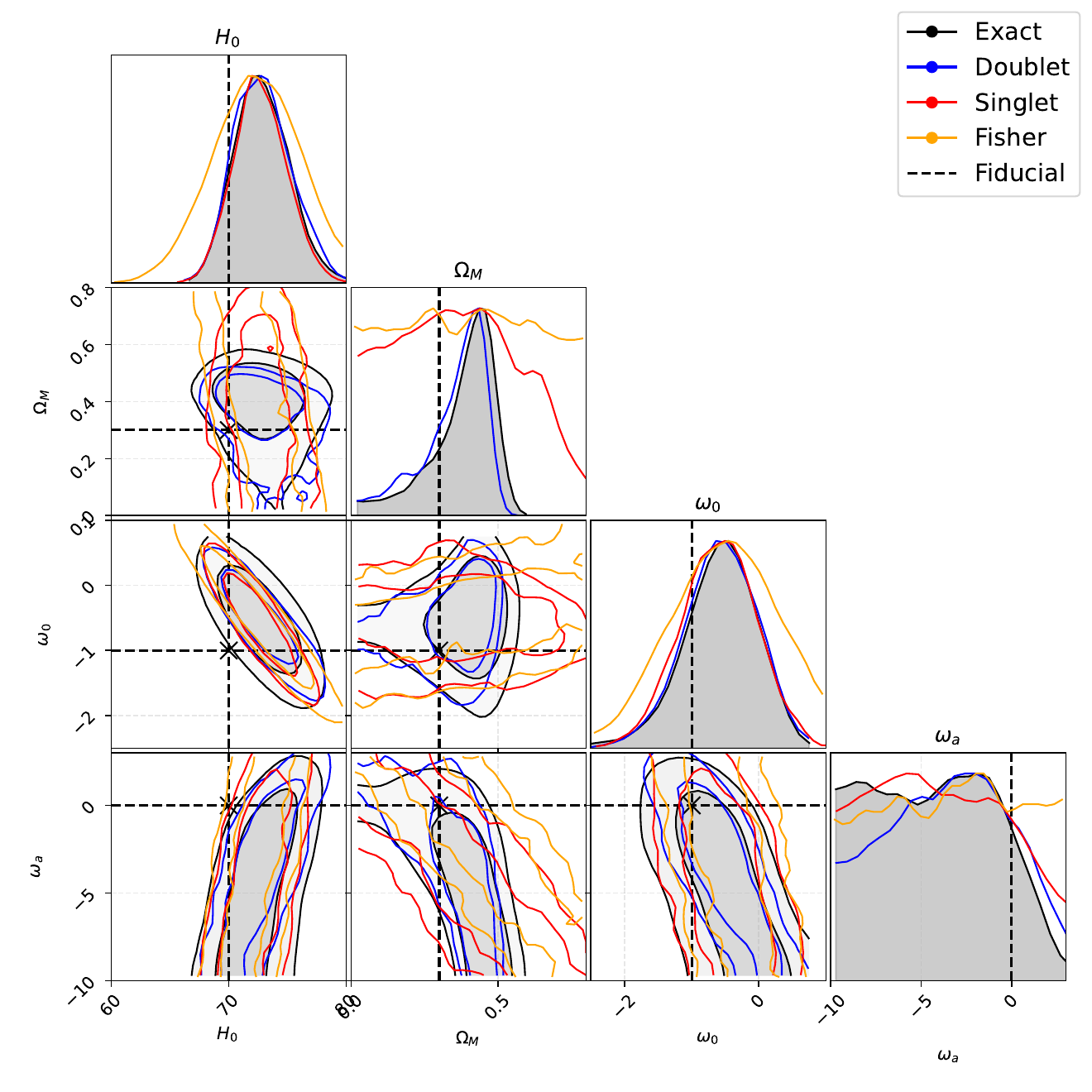}
    \caption{Cosmological corner plot from 300 GW sources (of a single catalog) up to $z=1.0$ comparing several approximations.}
    \label{fig:CornerCosmo0018}
\end{figure}

\subsection{Uncertainties, Biases and Divergences}
    
\hspace{1.0cm}Along this work we refer the bias as the displacement of some parameter measurement $\langle X\rangle$ from its fiducial value $X_{fid}$ weighted by the uncertainty $\sigma_X$ on that measurement (with a confidence level of $68.27\%$) $$ bias(X) \equiv \frac{|\langle X\rangle  - X_{fid}|}{\sigma_X}\, .$$

Following~\cite{josiel_2026} we quantify the comparisons among posteriors (exact against Fisher/DALI approximations) by using the statistical measurement called Jessen-Shannon Divergence (JSD) which quantifies how far some distribution $Q$ is from another distribution $P$  where mathematically is given by~\cite{Lin-JSD-1991}:
$$D_{JS}(P,Q) \equiv \frac{1}{2} \left[ D_{KL}\left(P\, \bigg\vert\, \frac{P+Q}{2}\right) + D_{KL}\left(Q\, \bigg\vert\, \frac{P+Q}{2}\right) \right] $$
which is defined in terms of the Kullback-Leibler divergence (KLD) defined bellow~\cite{kullback1951information}.
$$D_{KL}(P|Q) \equiv \int P(\mathbf{x})\cdot log\left(\frac{P(\mathbf{x})}{Q(\mathbf{x})}\right)d\mathbf{x}\, .$$

\subsection{Cosmological Results}

\hspace{1.0cm}We now compare the propagation of the three GW inference prescriptions, Fisher Singlet, Fisher with Gaussian Priors and Doublet, into the cosmological analysis. For each catalog construction, the cosmological likelihood is independently evaluated using the exact, Doublet, Singlet, and Fisher approximations. This distinction separates the approximation used to reconstruct the GW luminosity-distance information from the approximation subsequently used to infer the cosmological parameters.

The comparison therefore addresses two related effects: how the GW-level likelihood approximation modifies the information entering the Hubble diagram, and how the cosmological likelihood approximation propagates this information into constraints on $H_0$, $\Omega_m$, $\omega_0$, and $\omega_a$. Across the three GW catalog constructions, the cosmological Doublet approximation follows the exact inference more closely than the Singlet and Fisher approximations for most parameter--redshift combinations, as also quantified by the Jensen--Shannon divergences discussed above.

\begin{figure}[h]
    \centering
    \includegraphics[width=\linewidth]{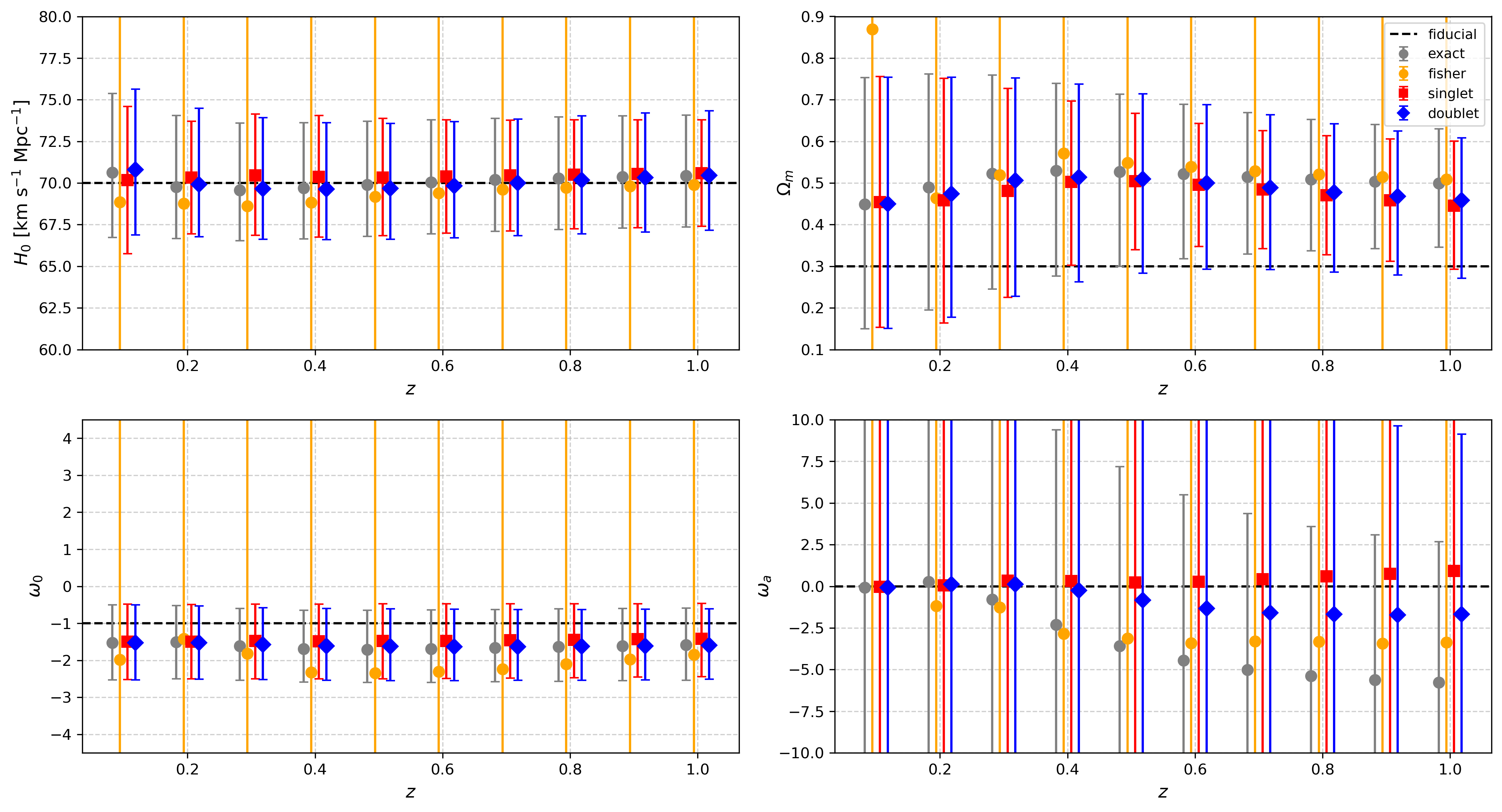}
    \caption{Cosmological constraints obtained from GW-Fisher Gaussian catalogs as a function of the maximum redshift included in the cumulative analysis. The panels show $H_0$, $\Omega_m$, $\omega_0$, and $\omega_a$ for the exact, Doublet, Singlet, and Fisher cosmological likelihoods.}
    \label{fig:cosmo_gwgaussian}
\end{figure}

\begin{figure}[h]
    \centering
    \includegraphics[width=\linewidth]{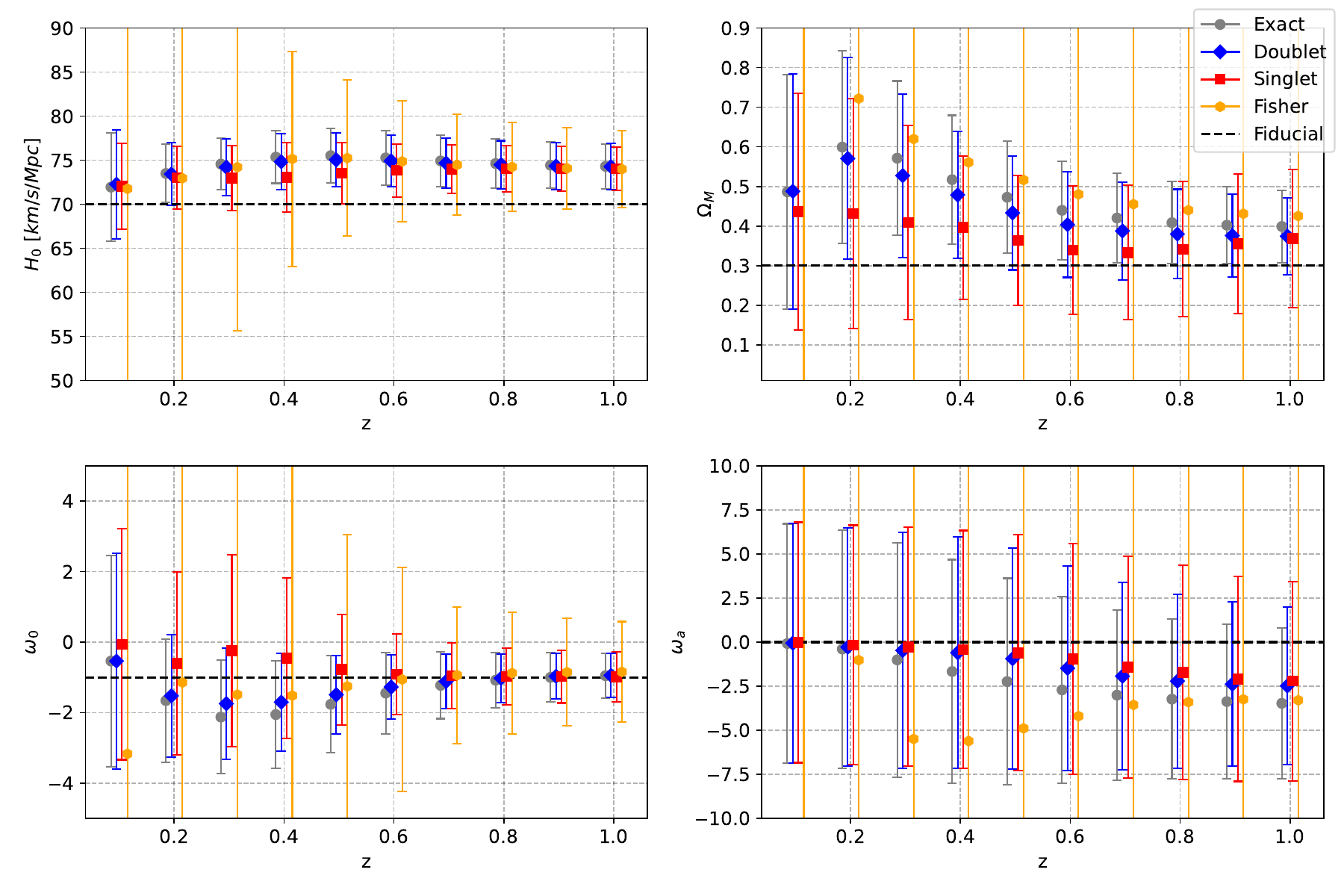}
    \caption{Same as Fig.~\ref{fig:cosmo_gwgaussian}, for GW-Singlet catalogs.}
    \label{fig:cosmo_gwsinglet}
\end{figure}

\begin{figure}[h]
    \centering
    \includegraphics[width=\linewidth]{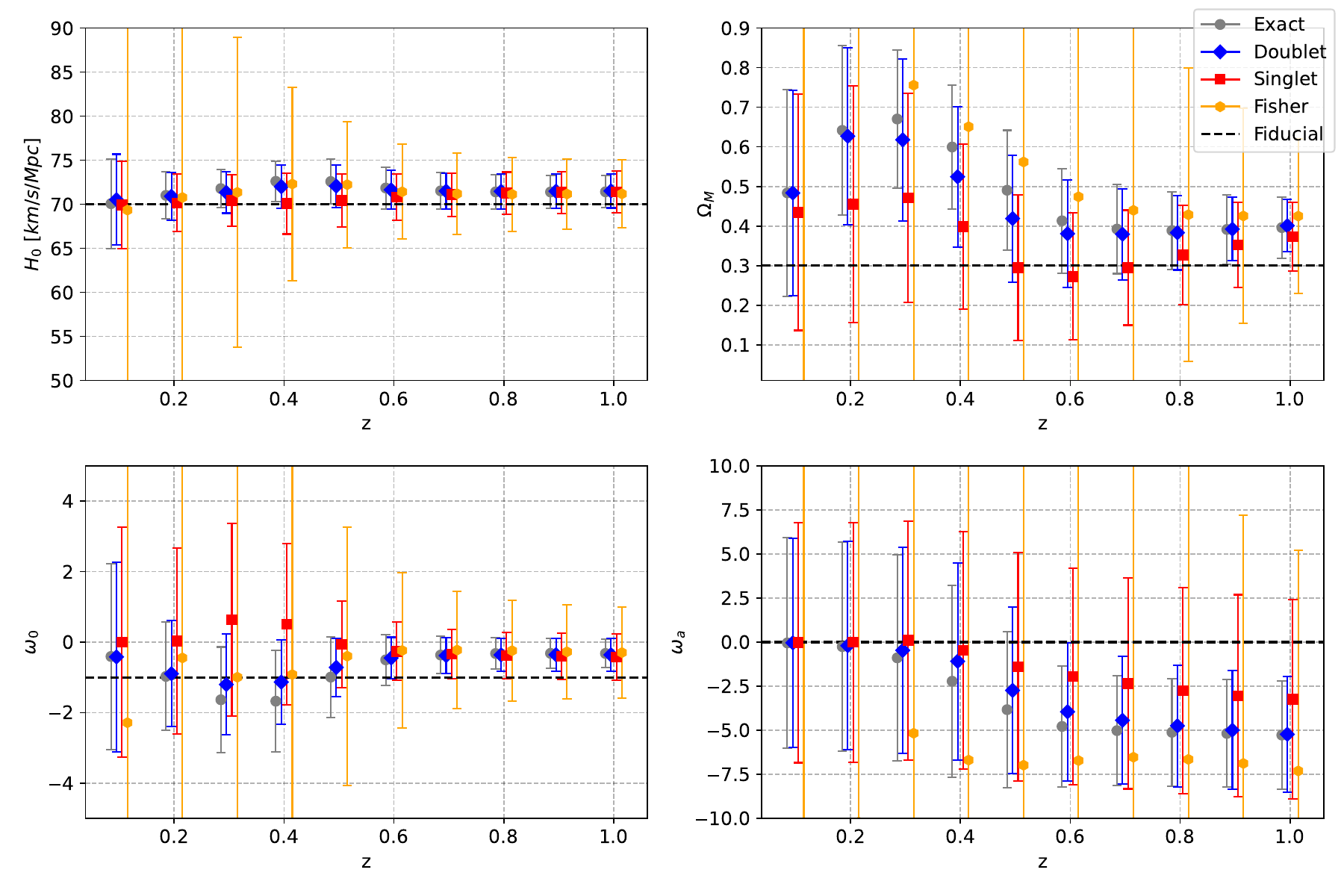}
    \caption{Same as Fig.~\ref{fig:cosmo_gwgaussian}, for GW-Doublet catalogs.}
    \label{fig:cosmo_gwdoublet}
\end{figure}
Figures~\ref{fig:cosmo_gwgaussian}--\ref{fig:cosmo_gwdoublet} show that the cosmological constraints depend on both the GW catalog construction and the likelihood approximation adopted at the cosmological stage. The differences among the GW-Fisher Gaussian, GW-Singlet, and GW-Doublet catalogs are propagated into the recovered central values and credible intervals, with the largest differences appearing for $\Omega_m$, $\omega_0$, and $\omega_a$. Within each catalog construction, however, the relative behavior of the cosmological approximations is similar: the Doublet reconstruction remains close to the exact result, the Singlet approximation shows larger departures in some redshift intervals, and the Fisher approximation generally produces broader credible intervals. Thus, although the GW-level approximation determines the information supplied to the cosmological analysis, the inclusion of second-order DALI terms at the cosmological level consistently reduces the discrepancy with respect to the exact likelihood.

\begin{figure}[h]
    \centering

    \begin{subfigure}[t]{0.32\linewidth}
        \centering
        \includegraphics[width=\linewidth]{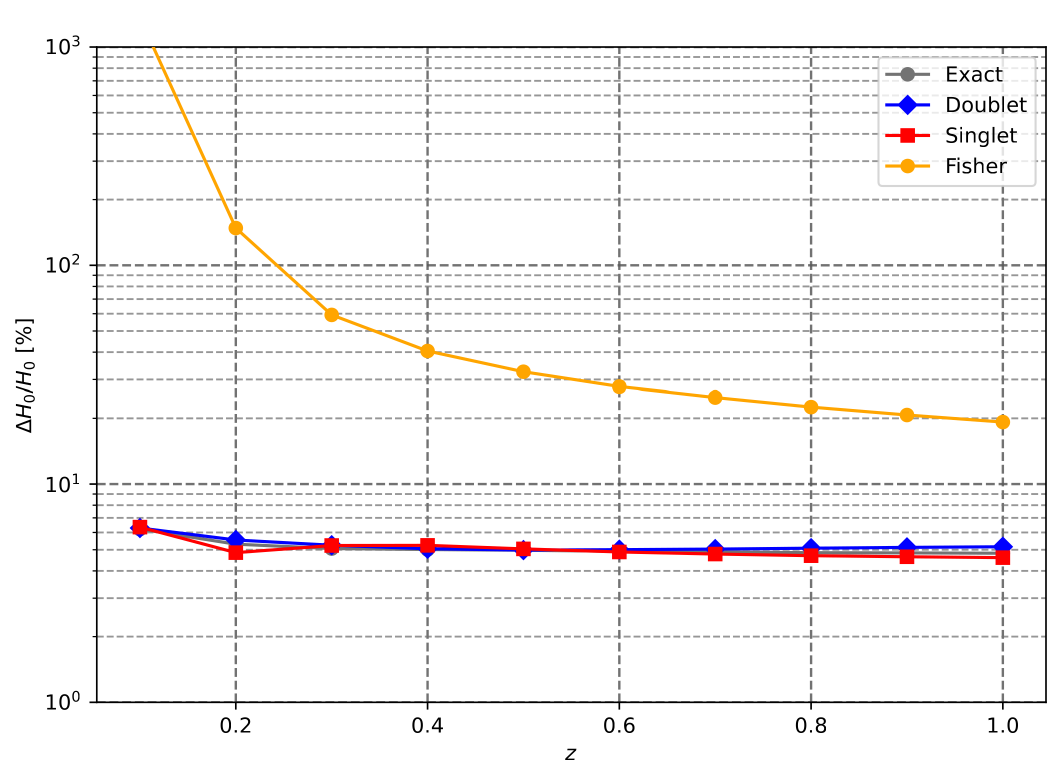}
        \caption{GW-Fisher Gaussian}
        \label{fig:h0_gwgaussian}
    \end{subfigure}
    \hfill
    \begin{subfigure}[t]{0.32\linewidth}
        \centering
        \includegraphics[width=\linewidth]{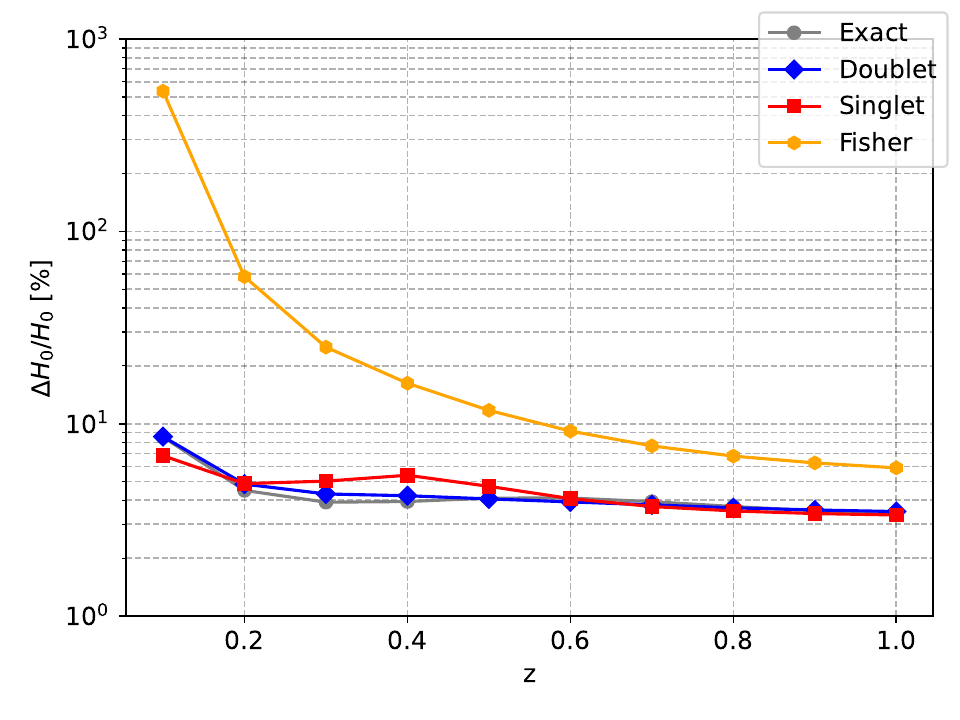}
        \caption{GW-Singlet}
        \label{fig:h0_gwsinglet}
    \end{subfigure}
    \hfill
    \begin{subfigure}[t]{0.32\linewidth}
        \centering
        \includegraphics[width=\linewidth]{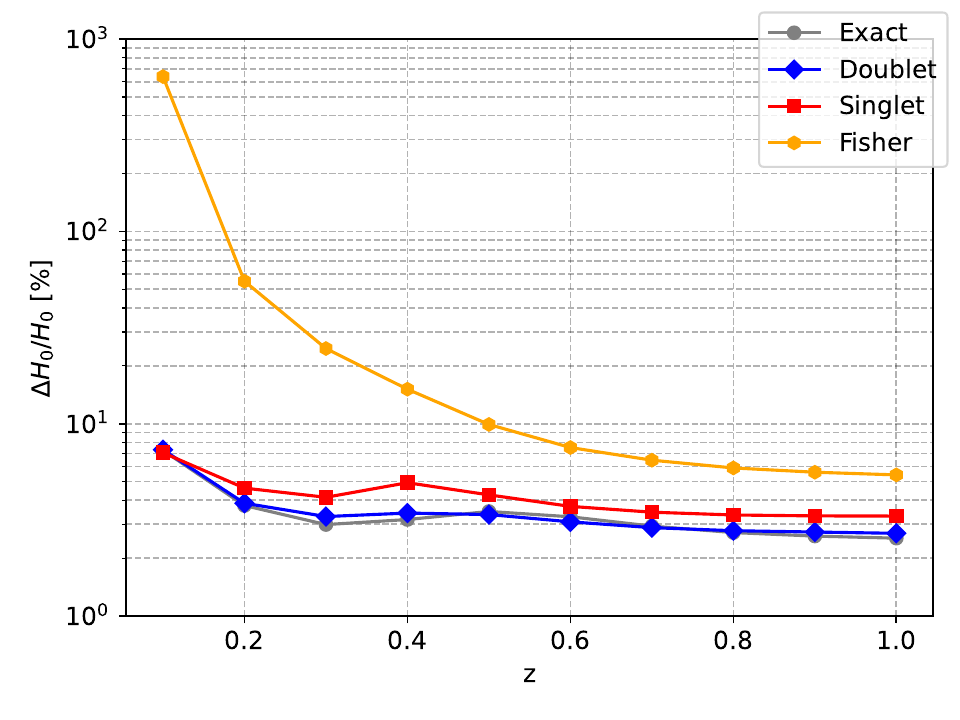}
        \caption{GW-Doublet}
        \label{fig:h0_gwdoublet}
    \end{subfigure}

    \caption{Relative uncertainty on the Hubble constant, $\Delta H_0/H_0$, as a function of the maximum redshift included in the cumulative analysis for the three GW catalog constructions. Each panel compares the exact, Doublet, Singlet, and Fisher cosmological likelihoods.}
    \label{fig:h0_catalog_comparison}
\end{figure}

The evolution of $\Delta H_0/H_0$ in Fig.~\ref{fig:h0_catalog_comparison} isolates the effect of the different approximations on the Hubble constant constraint. To compare the complete cosmological parameter set at the final cumulative redshift, Tables~\ref{tab:cosmo_gwgaussian}--\ref{tab:cosmo_gwdoublet} summarize the constraints obtained at $z_{\max}=1.0$. Focusing on $H_0$, the exact, Doublet, and Singlet approximations yield comparable statistical uncertainties, while the Fisher approximation consistently produces broader constraints across the three catalog configurations. The differences among the approximations are comparatively small in terms of $\Delta H_0/H_0$, indicating that the inferred precision on $H_0$ is relatively stable under these treatments. By contrast, the associated bias shows a stronger dependence on the adopted approximation and catalog construction. In particular, the Fisher approximation generally exhibits a smaller bias than the other methods, despite its larger uncertainty. These results indicate that the choice of approximation affects not only the width of the inferred $H_0$ constraints but also the displacement of the recovered value from the fiducial cosmology.

\begin{table}[h]
\centering
\setlength{\tabcolsep}{6pt}
\renewcommand{\arraystretch}{1.08}
\begin{tabular}{c | c c | c c | c c | c c  }
\hline
\textbf{Method} & \textbf{$\Delta H_0/H_0$} & $bias(H_0)$ & \textbf{$\Delta \Omega_M$} & $bias(\Omega_M)$ & \textbf{$\Delta \omega_0$} & $bias(\omega_0)$ & \textbf{$\Delta \omega_a$} & $bias(\omega_a)$ \\
\hline
Exact   & $4.82\%$ & $20.7\%$ & $0.14$ & $130.7\%$ & $0.98$ & $57.9\%$ & $8.91$ & $66.4\%$ \\
Doublet & $5.16\%$ & $20.6\%$ & $0.17$ & $88.0\%$ & $0.96$ & $59.3\%$ & $10.72$ & $14.2\%$\\
Singlet & $4.60\%$ & $18.7\%$ & $0.15$ & $104.1\%$ & $0.99$ & $44.9\%$ & $13.23$ & $5.0\%$\\
Fisher  & $19.23\%$ & $2.2\%$ & $0.67$ & $33.0\%$ & $6.63$ & $12.2\%$ & $101.74$ & $3.5\%$\\
\hline
\end{tabular}
\caption{{Cosmological constraints with GW-Fisher Gaussian catalogs up to $z=1.0$ (300 GW sources)}}
\label{tab:cosmo_gwgaussian}
\end{table}

\begin{table}[h]
\centering
\setlength{\tabcolsep}{6pt}
\renewcommand{\arraystretch}{1.08}
\begin{tabular}{c | c c | c c | c c | c c  }
\hline
\textbf{Method} & \textbf{$\Delta H_0/H_0$} & $bias(H_0)$ & \textbf{$\Delta \Omega_M$} & $bias(\Omega_M)$ & \textbf{$\Delta \omega_0$} & $bias(\omega_0)$ & \textbf{$\Delta \omega_a$} & $bias(\omega_a)$ \\ \hline
Exact & $3.40\%$ & $170.17\%$ & $0.092$ & $107.92\%$ & $0.634$ & $7.73\%$ & $4.290$ & $80.92\%$ \\
Doublet & $3.50\%$ & $164.52\%$ & $0.097$ & $76.28\%$ & $0.620$ & $8.96\%$ & $4.468$ & $55.51\%$ \\
Singlet & $3.36\%$ & $161.86\%$ & $0.174$ & $39.26\%$ & $0.710$ & $2.08\%$ & $5.655$ & $39.04\%$ \\
Fisher & $5.90\%$ & $91.20\%$ & $0.760$ & $16.48\%$ & $1.426$ & $11.06\%$ & $19.781$ & $16.66\%$ \\
\hline
\end{tabular}
\caption{{Cosmological constraints with GW-Fisher Singlet catalogs up to $z=1.0$ (300 GW sources)}}
\end{table}

\begin{table}[h]
\centering
\setlength{\tabcolsep}{6pt}
\renewcommand{\arraystretch}{1.08}
\begin{tabular}{c | c c | c c | c c | c c}
\hline
\textbf{Method}
& \textbf{$\Delta H_0/H_0$} & $\mathrm{bias}(H_0)$
& \textbf{$\Delta\Omega_m$} & $\mathrm{bias}(\Omega_m)$
& \textbf{$\Delta \omega_0$} & $\mathrm{bias}(\omega_0)$
& \textbf{$\Delta \omega_a$} & $\mathrm{bias}(\omega_a)$ \\
\hline
Exact   & $2.54\%$ & $77.8\%$ & $0.08$ & $123\%$ & $0.40$ & $169\%$ & $3.07$ & $172\%$ \\
Doublet & $2.69\%$ & $78.2\%$ & $0.07$ & $152\%$ & $0.46$ & $139\%$ & $3.29$ & $159\%$ \\
Singlet & $3.31\%$ & $59.9\%$ & $0.09$ & $84.3\%$ & $0.66$ & $87.7\%$ & $5.66$ & $53.3\%$ \\
Fisher  & $5.42\%$ & $30.5\%$ & $0.19$ & $64.1\%$ & $1.28$ & $54.9\%$ & $12.51$ & $58.4\%$ \\
\hline

\end{tabular}
\caption{Cosmological constraints with GW-Doublet catalogs up to $z_{\max}=1.0$ (300 GW sources).}
\label{tab:cosmo_gwdoublet}
\end{table}

\subsection{Jessen-Shannon Divergences Comparison}

\begin{figure}[h]
    \centering
    \includegraphics[width=\linewidth]{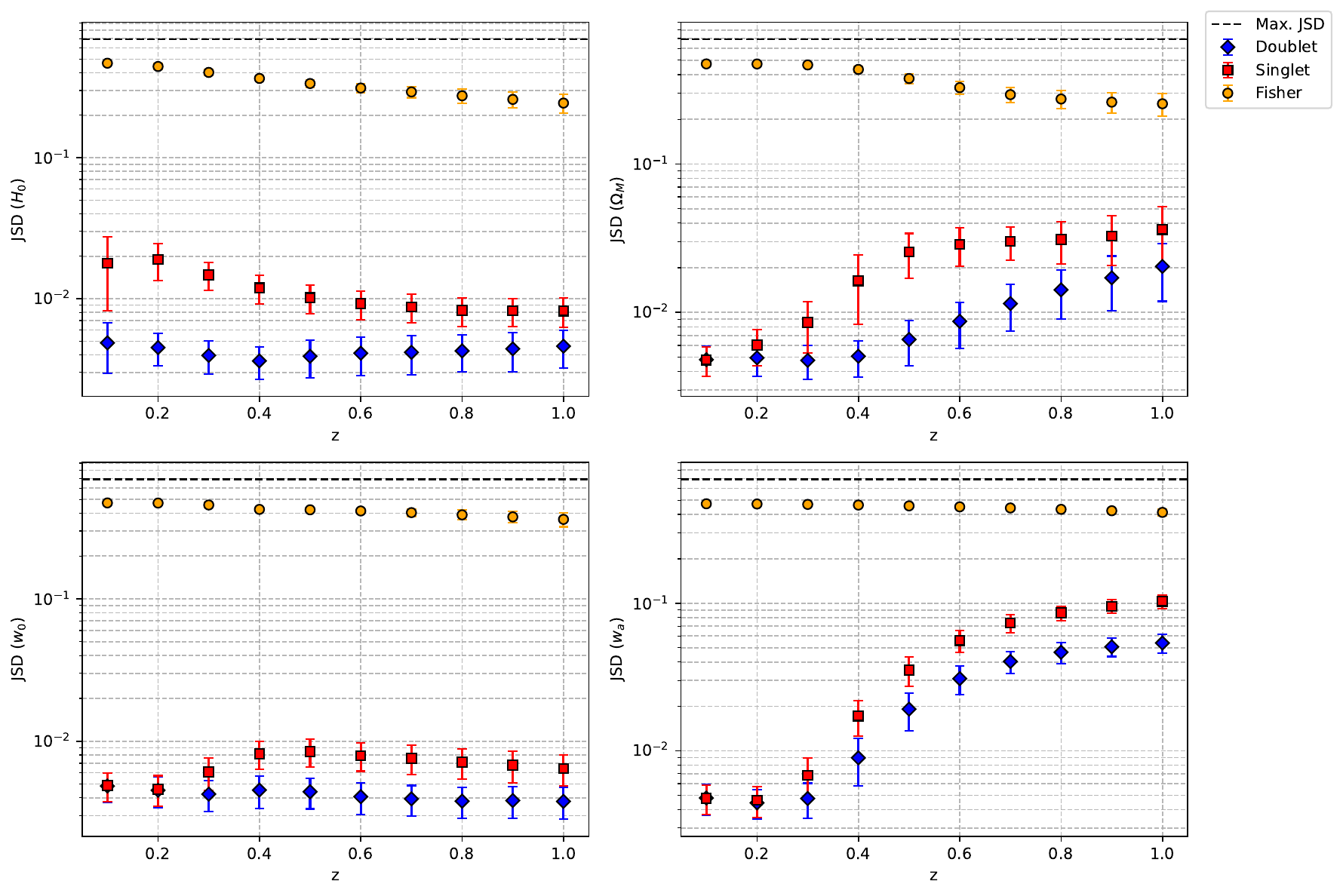}
    \caption{One-dimensional Jessen-Shannon Divergences for cosmological constraints using GW-Fisher Gaussian catalogs.}
    \label{Fig.JsdGW-Gaussian}
\end{figure}

\begin{figure}[h]
    \centering
    \includegraphics[width=\linewidth]{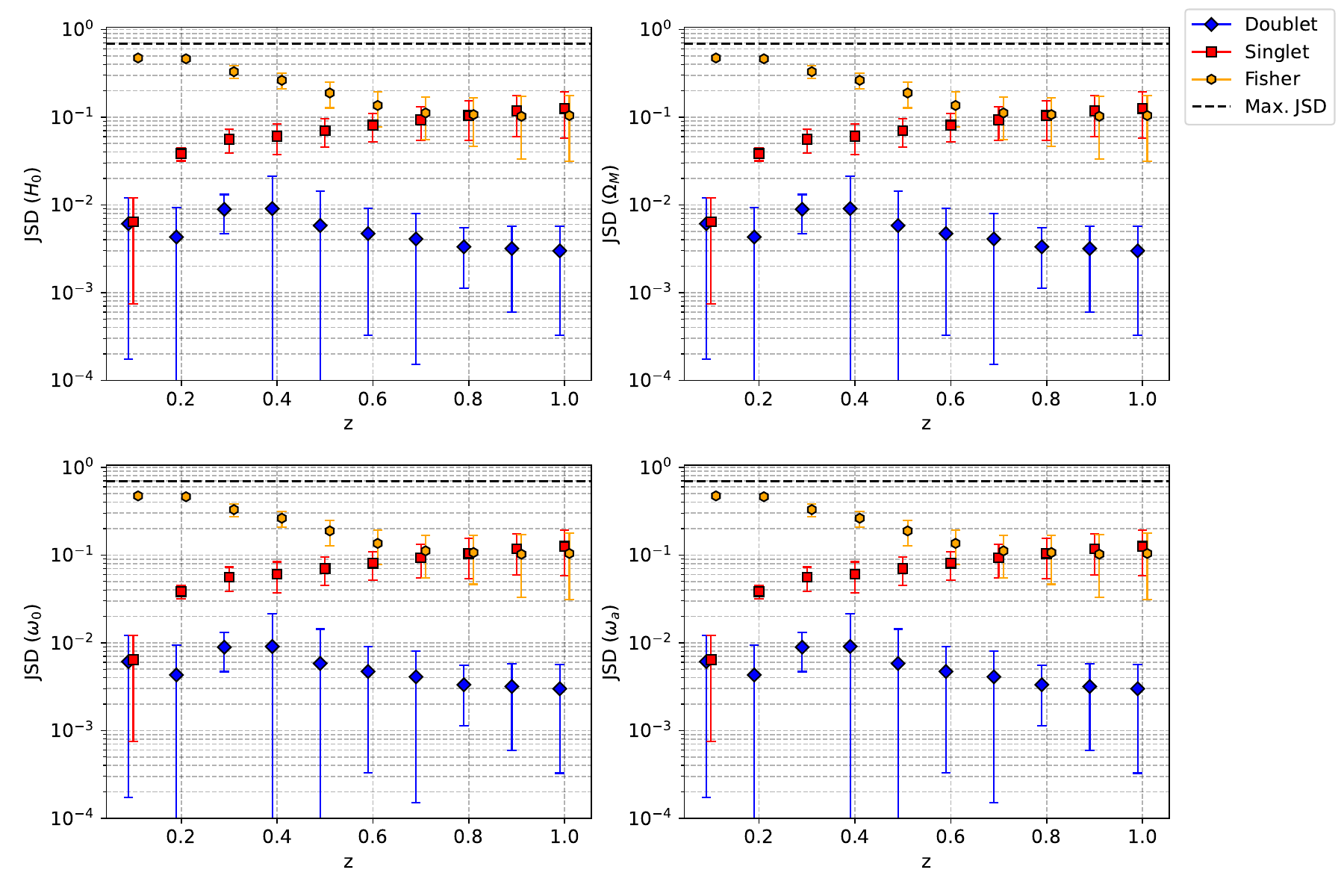}
    \caption{One-dimensional Jessen-Shannon Divergences for cosmological constraints using GW-Singlet catalogs.}
    \label{Fig. JsdGW_SInglet}
\end{figure}

\begin{figure}[h]
    \centering
    \includegraphics[width=\linewidth]{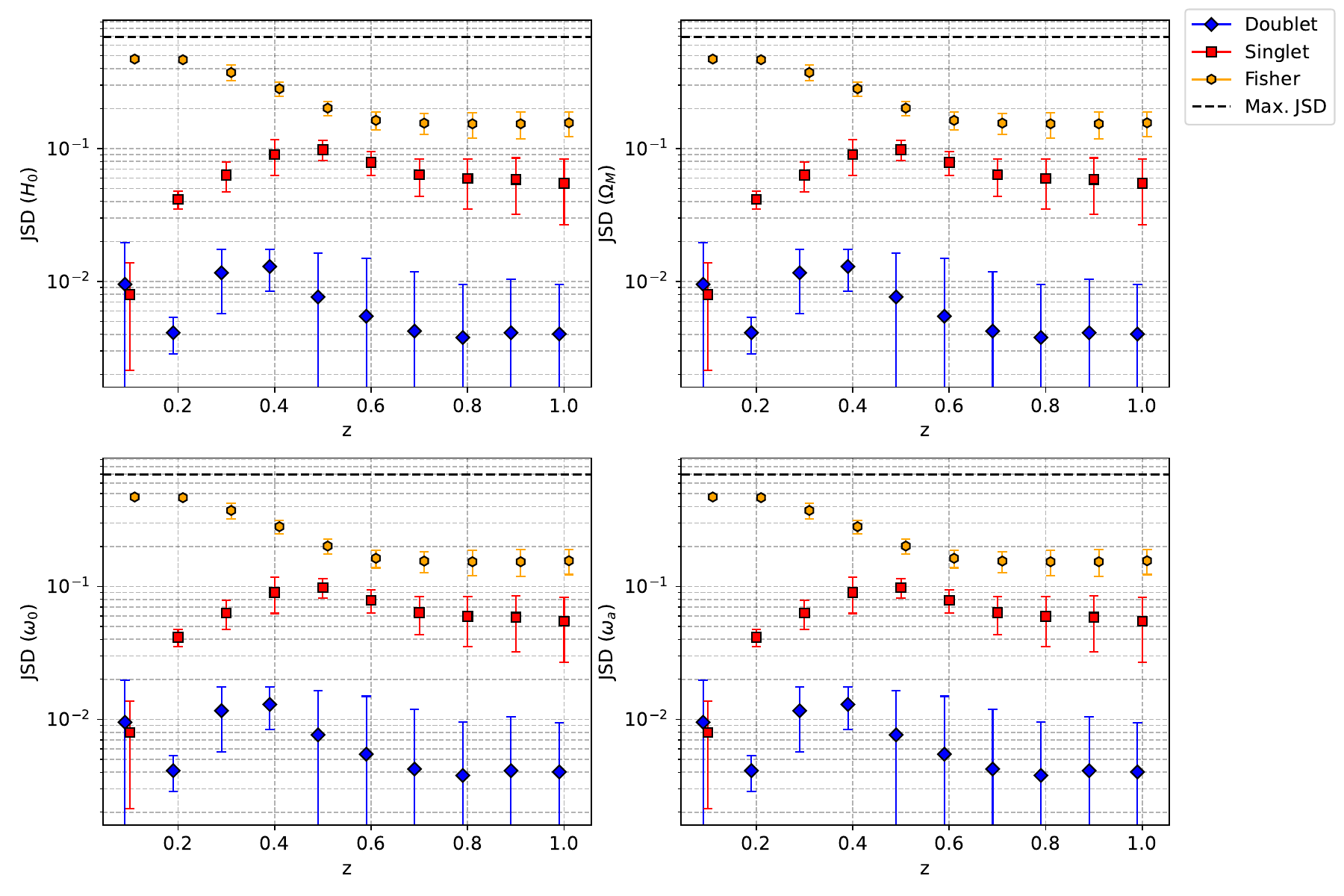}
    \caption{One-dimensional Jessen-Shannon Divergences for cosmological constraints using GW-Doublet catalogs.}
    \label{Fig.JsdGW-Doublet}
\end{figure}

\hspace{1.0cm}Across the three catalog constructions, the comparison shows a consistent ordering of the central JSD values, with the Doublet approximation yielding lower divergences than the Singlet and Fisher approximations for most cosmological parameters and redshift bins (Figs.~\ref{Fig.JsdGW-Gaussian}--\ref{Fig.JsdGW-Doublet}). The Singlet results generally occupy an intermediate range, while the Fisher approximation produces the largest divergences. This ordering is preserved despite differences in the redshift dependence among the cosmological parameters, indicating that the relative agreement of the three approximations with the exact posterior is primarily determined by the approximation scheme rather than by the catalog construction.

\color{black}

\subsection{Evaluation Time}
\label{sec:computational_performance}

\begin{figure}[h] 
\centering 
\includegraphics[width=.6\linewidth]{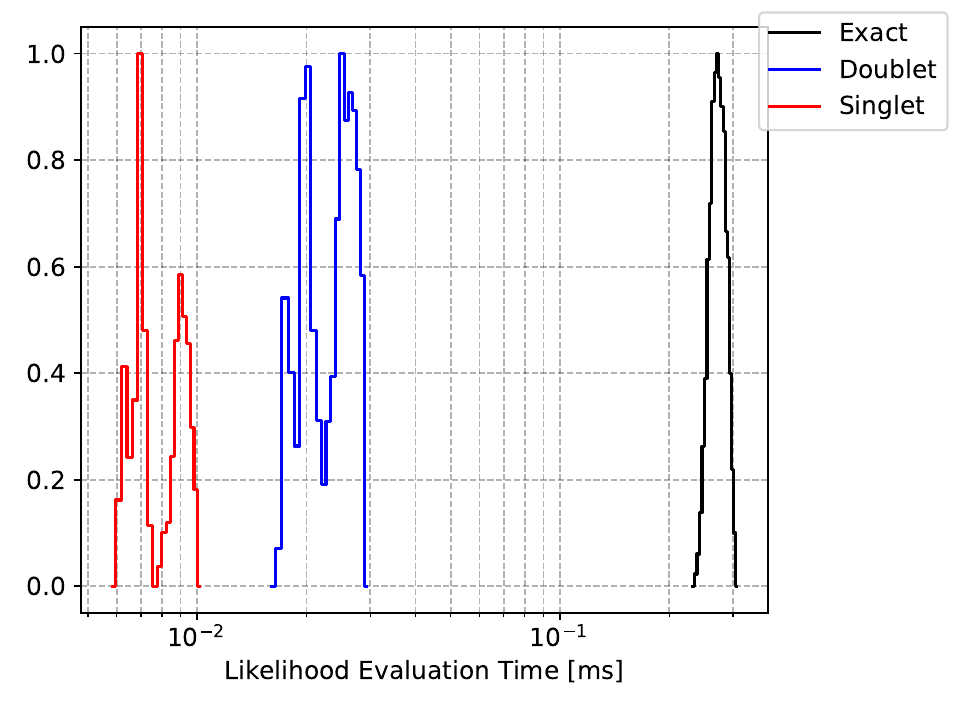} \caption{Distribution of cosmological likelihood evaluation times among different approximations.} 
\label{Fig.CosmoTime} 
\end{figure}

\hspace{1.0cm} In addition to the accuracy of the different likelihood approximations, their computational cost is relevant for cosmological inference, where the likelihood must be evaluated repeatedly during posterior sampling. Figure~\ref{Fig.CosmoTime} shows the distribution of the cosmological likelihood evaluation times for the exact, Doublet, and Singlet prescriptions, considering 80 catalogs containing 300 sources up to $z_{\max}=1.0$.

The mean evaluation time obtained for the exact likelihood is
$\bar{t}_{\mathrm{exact}}=2.7\times10^{-1}\,\mathrm{ms}$, compared with
$\bar{t}_{\mathrm{Doublet}}=2.3\times10^{-2}\,\mathrm{ms}$ and
$\bar{t}_{\mathrm{Singlet}}=8.0\times10^{-3}\,\mathrm{ms}$.
The Doublet approximation is therefore approximately 12 times faster than the exact likelihood, while the Singlet approximation reduces the evaluation time by a factor of approximately 34. The distributions in Fig.~\ref{Fig.CosmoTime} also show a clear separation among the three prescriptions, indicating that the reduction in computational cost is systematic across the analyzed catalogs.

 Although Singlet provides the lowest computational cost, the previous results show that the inclusion of the higher-order Doublet terms substantially improves the description of the non-Gaussian likelihood structure. Doublet therefore provides a favorable compromise in the present analysis: it retains part of the higher-order information beyond the Fisher approximation while remaining more than one order of magnitude faster to evaluate than the exact likelihood.
    
% <<<<<<<<<<<<<<<<<<<<<<<<<<<<<<<<<<<<<<<<<<<<<<<<<<<<<<<<<<<<<<<<<<<<<<<<<<<<<<<<<<<<<<<<<<<<<<<<<<<<<<<<<<<<<<<<<<<<<<<<<<<

% <<<<<<<<<<<<<<<<<<<<<<<<<<<<<<<<<<<<<<<<<<<<<<<<<<<<<<<<<<<<<<<<<<<<<<<<<<<<<<<<<<<<<<<<<<<<<<<<<<<<<<<<<<<<<<<<<<<<<<<<<<<

% <<<<<<<<<<<<<<<<<<<<<<<<<<<<<<<<<<<<<<<<<<<<<<<<<<<<<<<<<<<<<<<<<<<<<<<<<<<<<<<<<<<<<<<<<<<<<<<<<<<<<<<<<<<<<<<<<<<<<<<<<<<

\section{Conclusions and perspectives}
\label{sec:end}

\hspace{1.0cm}In this paper, we have conducted a systematic analysis of the impact of different-quality bright siren simulations, using Gaussian and beyond-Gaussian likelihood approximations to constrain cosmological parameters. We found that, by using the best Hubble diagram approximation (via the GW-Doublet catalogue), the uncertainties in the cosmological parameters are reduced by approximately a factor of two, leading to tighter constraints. However, we also find that the bias in the cosmological parameters can be larger for the Doublet reconstruction, by factors of approximately \(\sim 2.4\)--\(2.7\) relative to the Fisher reconstruction, depending on the parameter. This is explained by the fact that the use of traditional priors in the GW analysis (particularly on $d_L$) can shift the measurements of $d_L$ away from the fiducial injections, which consequently increases the bias in the cosmological measurements. It is naturally expected that the use of Gaussian priors (centered on the injections) on GW parameters in the Doublet approximation can significantly reduce the bias in the cosmological parameters, leading to results similar to those obtained using Fisher with Gaussian priors. However, in this work, our goal was to investigate the performance of the DALI approximation as a computationally efficient representation of the traditional GW simulations based on exact likelihoods and traditional priors on the GW parameters.

Given the several GW approaches for Hubble diagrams, we also investigated the ability of the cosmological DALI formalism to recover results similar to those obtained using the exact cosmological likelihoods by comparing parameter uncertainties and Jensen--Shannon divergences. As expected, based on the Jensen–Shannon divergence, the Doublet approximation provides the closest approximation to the reference posterior among the methods considered. Within the Fisher-based approaches, Fisher sampling (Singlet-DALI) generally provides a closer approximation than the direct Fisher reconstruction. For the GW-Doublet catalogue, the cosmological uncertainties obtained with the Doublet likelihood differ from the exact results by only $\sim 10\%$ on average across the four cosmological parameters, compared with $\sim 195\%$ when using the traditional Fisher approximation. Therefore, the cosmological Doublet approximation is, on average, approximately $19$ times closer to the exact likelihood results than the Fisher approximation.

Finally, we also investigated the computational cost of the different approximations at both the cosmological and GW levels. For the cosmological analysis, the Doublet likelihood evaluation is approximately 12 times faster than the exact likelihood, with mean evaluation times of $2.3\times10^{-2}$ ms and $2.7\times10^{-1}$ ms, respectively. To estimate the total computational cost of the combined GW and cosmological analyses, we further assume, as an approximate scaling, that the number of likelihood evaluations is proportional to the dimensionality of the parameter space. Since the GW analysis here involves six free parameters, compared with four for the cosmological analysis, we therefore take $N_{\rm eval}^{\rm cosmo}/N_{\rm eval}^{\rm GW}\simeq4/6$. Combining this assumption with the GW computational costs reported in previous work~\cite{josiel_2026}, where the Doublet approximation reduces the total GW MCMC cost by a factor of $\sim55$ relative to the exact likelihood, we estimate that the total computational cost of the combined GW and cosmological analysis is reduced by approximately the same factor of $\sim55$. This is because the GW analysis dominates the total computational cost, while the cosmological Doublet provides an additional $\sim12\times$ reduction in the cost per likelihood evaluation. These estimates should be regarded as approximate, since the actual number of likelihood evaluations depends on the sampling algorithm and convergence criteria.

Overall, this work demonstrates the potential of DALI-based approximations to provide accurate and computationally efficient tools forforecasts in bright siren cosmology, while also highlighting the importance of carefully accounting for biases introduced by the underlying GW analysis. These results also motivate the exploration of similar approximation and optimization strategies in the context of dark sirens, where the cosmological likelihoods involve additional information from the galaxy distribution and can be substantially more computationally demanding. In this context, the use of efficient likelihood approximations and other computational optimization strategies may become particularly relevant for enabling cosmological inference with the large number of GW events expected from future detectors. Future work could extend this analysis to state-of-the-art waveform
models for binary neutron stars that incorporate tidal effects and
neutron-star tidal deformabilities, such as TEOBResumS~\cite{Nagar:2018zoe}
and waveform models incorporating the NRTidalv3 prescription
~\cite{Abac2024}, and investigate the impact of updated detector
sensitivity curves on the resulting parameter-estimation and
cosmological forecasts.

\acknowledgments
MHP is supported by the Brazilian agency \textit{Conselho Nacional de Desenvolvimento Científico e Tecnológico} (CNPq). JMSdS is supported by the Brazilian research agency \textit{Fundação de Amparo à Pesquisa e Inovação do Estado do Espírito Santo} (FAPES). He also thanks \textit{Fundação Carlos Chagas Filho de Amparo à Pesquisa do Estado do Rio de Janeiro} (FAPERJ), process number E-26/200.236/2024 and E-26/200.237/2024. ODA thanks the Brazilian Ministry of Science, Technology
and Innovation and the Brazilian Space Agency (AEB), which supported the present work under PO 20VB.0009. He also acknowledges CNPq for the \textit{Bolsa de Produtividade em Pesquisa} (PQ), process number 310087/2021-0. We acknowledge the use of the computational resources of the joint CHE / Milliways cluster, supported by a FAPERJ grant E-26/210.130/2023. The authors also thank to MCTIC/FINEP (CT-INFRA grant 0112052700) and the Embrace Space Weather Program for the computing facilities at INPE.

%\color{blue}
%\section*{Appendix}
%\appendix

% ---- Bibliography (BibTeX) ----
\bibliographystyle{JHEP}
\bibliography{biblio}

\end{document}